\documentclass[fleqn,usenatbib]{mnras}

\usepackage[T1]{fontenc}

\DeclareRobustCommand{\VAN}[3]{#2}
\let\VANthebibliography\thebibliography
\def\thebibliography{\DeclareRobustCommand{\VAN}[3]{##3}\VANthebibliography}

\usepackage{graphicx}	
\usepackage{amsmath}	
\usepackage{amssymb}	

\title[Stability around Saturnian small moons]{Dynamical stability in the vicinity of Saturnian small moons. The cases of Aegaeon, Methone, Anthe and Pallene}

\author[A. Rodr\'iguez et al.]{
Adri\'an Rodr\'iguez,$^{1}$\thanks{E-mail: adrian@astro.ufrj.br}
N. Callegari Jr.$^{2}$
\\
$^{1}$Observat\'orio do Valongo, Universidade Federal do Rio de Janeiro, Ladeira do Pedro Ant\^ onio 43, 20080-090, Rio de Janeiro, Brazil\\
$^{2}$S\~ao Paulo State University (UNESP), Institute of Geosciences and Exact Sciences, Av. 24-A, 1515, 13506-900, Rio Claro, SP, Brazil
}

\date{Accepted XXX. Received YYY; in original form ZZZ}

\pubyear{2021}

\begin{document}
\label{firstpage}
\pagerange{\pageref{firstpage}--\pageref{lastpage}}
\maketitle


\begin{abstract}
In this work we analyze the orbital evolution and the dynamical stability in the vicinity of the small saturnian moons Aegaeon, Methone, Anthe and Pallene. We numerically resolve the exact equations of motions to investigate the orbital motion of thousands of test particles within and near to the domain of the 7/6, 14/15, 10/11 mean motion resonances of Aegaeon, Methone and Anthe with Mimas, respectively. We show that, for massless small moons, the orbits of particles initially restricted to the resonant domains remain stable for at least $10^4$ yr. We also conduct numerical simulations considering Aegaeon, Methone, Anthe and Pallene as massive bodies. The results show that most particles undergo significant perturbations in their orbital motions, ultimately destabilizing in timescales of a few hundreds of years or even less through collisions with the four small moons. In addition, we also simulate the orbital evolution of test particles initially distributed in form of arc around Aegaeon, Methone and Anthe. We show that the initial arcs are dynamically eroded on timescales of hundreds of years, allowing us to constraint the timescales for which gravitational forces operate to remove particles from the observed arcs.
\end{abstract}




\section{Introduction}
\label{intro}

\onecolumn

The saturnian system is composed by a collection of mid-sized (few hundreds of km) and small (few km) satellites, being Titan the only large body (thousands of km) orbiting Saturn. Among the small satellites, Aegaeon, Methone, Anthe and Pallene (hereafter, small moons) have been observed by the Cassini mission between 2004 and 2009 \citep{porco+2005,spitale+2006,jacobson+2006,jacobson+2008,cooper+2008,hedman+2010}. They are irregularly shaped satellites with radii ranging between roughly  0.5 and 2.5 km \citep{thomas+2020} and occupying a region from 2.78 to 3.5 $R_S$, where $R_S$ is the radius of Saturn (see Table \ref{tabela-small-moons}). Aegaeon, Methone and Anthe orbit Saturn in resonant motion with Mimas, the closest mid-sized satellite. Specifically, the first order mean motion ressonances (MMR) are 7/6, 14/15 and 10/11 between Aegaeon, Methone and Anthe with Mimas, respectively \citep{spitale+2006,cooper+2008,hedman+2010}. Moreover, the resonant motions of which these satellites are trapped in are such that the librating critical angles involve the longitude of pericentre of the perturbing body, that is, Mimas. This particular orbital configuration, named corotation eccentricity resonance (hereafter, CER), allows the satellites to maintain very small values of their eccentricities (see Table \ref{tabela-small-moons}). 

Images taken by \textit{Cassini} also revealed the presence of arcs structures around Aegaeon, Methone and Anthe.  The arcs are populated by $\mu$m to cm-size dust grains particles and their origins are unclear, although collisions of micrometeoroids with the small moons can be a possibility. In fact, due to their small gravity, small satellites (5 km to 50 km) can be important sources for supplying debris to a ring or arc through collisions with micrometeoroids, since they are unable to retain the large amount of ejected material \citep{burns+1999}. The arcs sizes varies in radial and longitudinal extent and the corresponding small moons reside within each arc, librating in their respective corotation resonances. According to \citet{hedman+2009}, the longitudinal sizes of the arcs are $\sim10^{\circ}$ and $\sim20^{\circ}$ for Methone and Anthe, respectively, whereas the arc of Aegaeon extends over $\sim60^{\circ}$ in longitude \citep{hedman+2007}. Previous works suggested that the CERs act to confine the particles of the arcs in longitudinal extent \citep[e.g.][]{hedman+2009}. However, the consideration of non-gravitational forces have shown that most particles are removed from the arcs in timescales of dozens of years, depending on the size of the grain \citep{sun+2017,madeira+2018,madeira+winter2020}, indicating that the origin and sustainability of the arcs is still a puzzle \citep[see also][]{araujo+2016}.

Some works have investigated the short and long-term dynamical evolution of Aegaeon, Methone, Anthe and Pallene \citep[e.g.][]{callegari+2010,callegari2020,munhoz+2017,callegari+2021}. However, the long-term orbital stability within and in the close neighborhood of the corotation resonances has been marginally investigated. Indeed, \citet{munhoz+2017} performed a dynamical analysis of the motion of test particles around Mimas under perturbations of Saturn's oblateness and companion satellites. They consider a wide interval of semi-major axes and do not focus in the dynamical stability within the CER regions. Moreover, their numerical simulations cover 18 yr of evolution. In this work we aim to analyze, by performing a large set of long-term numerical simulations with thousands of test particles, the orbital stability of massless bodies in the close vicinity of Aegaeon, Methone, Anthe and Pallene. In addition, we also aim to numerically explore the dynamical environment of those regions associated to the arcs of particles of the resonant moons. Despite that dust particles are subjected to non-gravitational forces, the purely conservative analysis allows us to constraint the timescales for which gravitational forces operate to remove particles from the three observed arcs.

Our model includes the gravitational perturbation of all mid-sized satellites, Titan and the contribution of the non-sphericity of Saturn ($J_2$ and $J_4$), solving the exact equations of motions for timescales of hundreds and thousands of years. We use several indicators of orbital motion such as the lifetime of the test particles and the maximum variations of semi-major axis, eccentricity and inclination. In the case of resonant small moons, we also investigate the dynamical survival of a distribution of test particles in arc-shaped initial configurations.

The paper is organized as follows: in Sec. \ref{model} we introduce the model adopted to perform the numerical simulations. The long-term orbital motion of test particles in the close vicinity of Aegaeon, Methone, Anthe and Pallene is addressed in Sec. \ref{massless}, considering the four small moons as massless bodies. We include the regions corresponding to the corotation resonances with Mimas, allowing us to highlight the domains and the dynamical behaviour within these resonances. In Sec. \ref{massive} we consider the four small moons as massive bodies and analyze the impact on the motion of surrounding particles within and in the vicinity of the corotation resonances. In addition, we investigate the orbital stability of arc-shaped initial configurations of particles around Aegaeon, Methone and Anthe, comparing the results with the observed arcs. Discussion and conclusions are devoted to Sec. \ref{discussion}.

\begin{table}
\begin{center}
\caption{Mean osculating orbital semi-major axis, eccentricity and inclination of the small moons. The orbital evolution follows the model described in Sec. \ref{model} and the initial values of orbital elements have been taken from \textit{Horizons} data base with date 2020/03/25.}
\vspace*{0.3cm}
\begin{tabular}{c | c c c}
\hline
Satellite & $a$ (km) & $e$ & $i$ (deg) \\ 
\hline
Aegaeon & 168,033 & 0.0032 & 0.001 \\
Methone & 194,698 & 0.0025 & 0.015 \\
Anthe & 198,106 & 0.0025 & 0.014 \\
Pallene & 212,705 & 0.0042 & 0.176\\
\hline
\label{tabela-small-moons}
\end{tabular}
\end{center}
\end{table}



\section{The model}
\label{model}

We consider the system composed by Saturn as the central mass orbited by Mimas, Enceladus, Tethys, Dione, Rhea, Titan, the small satellites Aegaeon,
Methone, Anthe, Pallene and a set of test massless particles. Initially, the test particles are distributed in the vicinity of each small moon, varying the initial values of semi-major axis and eccentricity. We assume that all orbital elements of test particles other than semi-major axis ($a$) and eccentricity ($e$), namely, the inclination ($i$), the argument of pericentre ($\omega$), the longitude of ascending node ($\Omega$) and the mean anomaly ($M$), are initially equal to the angles of each small moon at the adopted date. This choice allows for particles start within or in the vicinity of mean-motion resonances between Aegaeon, Methone and Anthe with Mimas \citep[see][]{callegari2020,callegari+2021}. {In addition, we also run a set of numerical simulations considering initial random angles ($\omega$ and $M$) for the test particles. This allows us to investigate the dynamical stability around other corotation sites which are not populated by dust particles.} For the current position of the small moons, the Sun's gravitational perturbation can be safely ignored and the dynamics is well described by the Saturn's gravitational field \citep[see Eqs. 4 and 5 of][]{callegari2001}.

We numerically integrate the exact equations of motions of the $N$-body problem by assuming mutual perturbations between all massive satellites and the contribution of the oblateness of Saturn through the $J_2$ and $J_4$ coefficients of the gravitational potential. We use the MERCURY integrator package \citep{chambers1999} choosing the Bulirsch-Stoer algorithm with a step-size of $1/20$ of the smallest orbital period of the system\footnote{{We also run considering a step-size of 1/80 of the smallest orbital period, however, the results are virtually the same.}}. For Saturn the mass and radius are, respectively, $M_S=5.68317\times10^{26}$ kg and $R_S=60,268$ km, whereas the $J_i$ coefficients are $J_2=1.629\times10^{-2}$ and $J_4=-9.337\times10^{-4}$ \citep{jacobson+2006}\footnote{We mention that \cite{Iess+2019} provide new measurements of the gravity filed of Saturn from the final phase of the \textit{Cassini} mission, improving the accuracy of the higher-order coefficients.}. The initial values of the osculating orbital elements have been taken from the \textit{Horizons} data base\footnote{https://ssd.jpl.nasa.gov/horizons.cgi}, with the date of 2020/03/25 for all satellites. The physical parameters of the mid-sized (and Titan) satellites have been also taken from the \textit{Horizons} data base.


\section{Numerical simulations with massless satellites}
\label{massless}

In this section we assume that Aegaeon, Methone, Anthe and Pallene are massless bodies, that is, there is no mutual interaction between them and also with test particles.
We numerically integrate the exact equations of motions according to the above described model considering a set of $4,900$ test particles
distributed along the $(a,e)$ space in the vicinity of each small moon, perturbed by the mid-sized satellites, Titan and for the gravitational potential of Saturn. The integration time adopted in each individual simulation is $10^4$ yr, representing roughly $\simeq4\times10^6$ orbital periods of the small moons. This integration time allows us to better characterize the long-term dynamical stability around the current resonant regions of the small moons with Mimas. We use the survival time of test particles ($\tau$) as the main indicator of orbital stability, however, for sake of completeness, we also show the maximum variations of semi-major axis ($\Delta a$), eccentricity ($\Delta e$) and inclination\footnote{The orbital inclination is here defined as the angle formed between the orbital plane and the equator of the central body at the considered epoch.} ($\Delta i$) to better understand the dynamical behaviour in the vicinity of the four small moons analyzed in this work. We define survivor as a test particle that does not results in ejection or collision with other bodies within the timespan adopted in this section.


In what follows we describe the results for each individual small moon.

\subsection{Methone}
\label{methone}

\begin{figure}
\begin{center}
\includegraphics[width=0.55\columnwidth,angle=270]{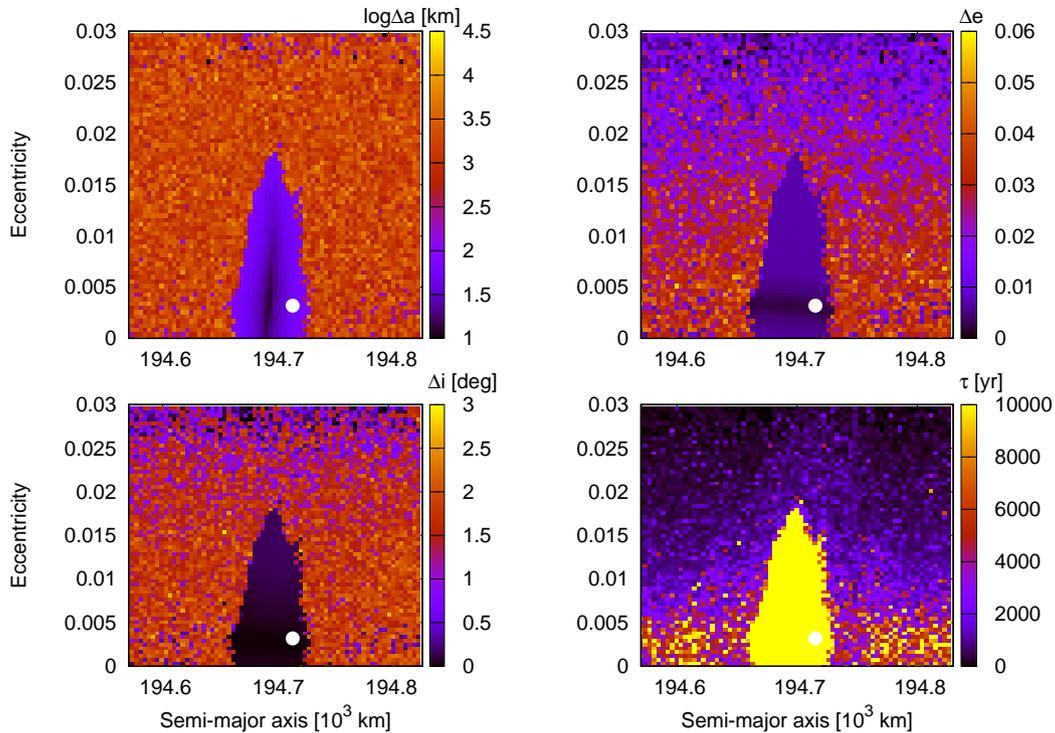}
\caption{ Numerical integration of 4,900 test particles with different initial values of semi-major axis and eccentricity in the vicinity of Methone, according to the model described in Sec. \ref{model}. The integration time is $10^4$ yr. For each particle we show the logarithm of the maximum attained variation of semi-major axis, $\log\Delta a$ (top left), the maximum variation of eccentricity, $\Delta e$ (top right), the maximum variation of inclination, $\Delta i$ (bottom left) and the survival time, $\tau$ (bottom right). The structure of the 14/15 corotation resonance with Mimas is easily seen in all panels. The filled dot indicates the values of Methone for the adopted date.}
\label{mapa-methone}
\end{center}
\end{figure}

Fig. \ref{mapa-methone} shows the result of the dynamical evolution of 4,900 test particles near to Methone. The lateral bar indicates the values of $\log\Delta a$, $\Delta e$, $\Delta i$ and $\tau$. The initial values of semi-major axis and eccentricity of particles range in 194,570 $\le a \le$ 194,830 km and\footnote{We have also extended to $e=0.07$ but none other structure other than the CER was found.} 0$\le e \le$ 0.03, respectively. The filled white circle shows the values of Methone for the adopted initial date. We clearly see the corotation zone associated to the external 14/15 MMR with Mimas in all panels, closely between 194,660 km and 194,730 km and $e<0.019$ \citep[see also][]{callegari+2021}. In this region, the critical angle $\sigma_C=15\lambda-14\lambda_{Mimas}-\varpi_{Mimas}$ librates around $0^{\circ}$, where $\lambda$ is the mean longitude of the test particle. We note that those particles within the CER zone survives for the whole integration time. In fact, surviving particles are mostly confined to the CER zone (see right bottom panel of Fig. \ref{mapa-methone}). The number of initial conditions which survive for $10^4$ yr is 628, representing $12.8\%$ of the total sample.  It is worth note that the result for $\Delta e$ (right top panel in Fig. \ref{mapa-methone}) shows an horizontal structure on the bottom part of the resonant region, which is characterized by small values of $\Delta e$ and thus indicating a very stable region. Moreover, Methone, identified with a white dot, is located inside of this structure. In addition, the width of the resonance decreases as the eccentricity increases.

\begin{figure}
\begin{center}
\includegraphics[width=0.6\columnwidth,angle=270]{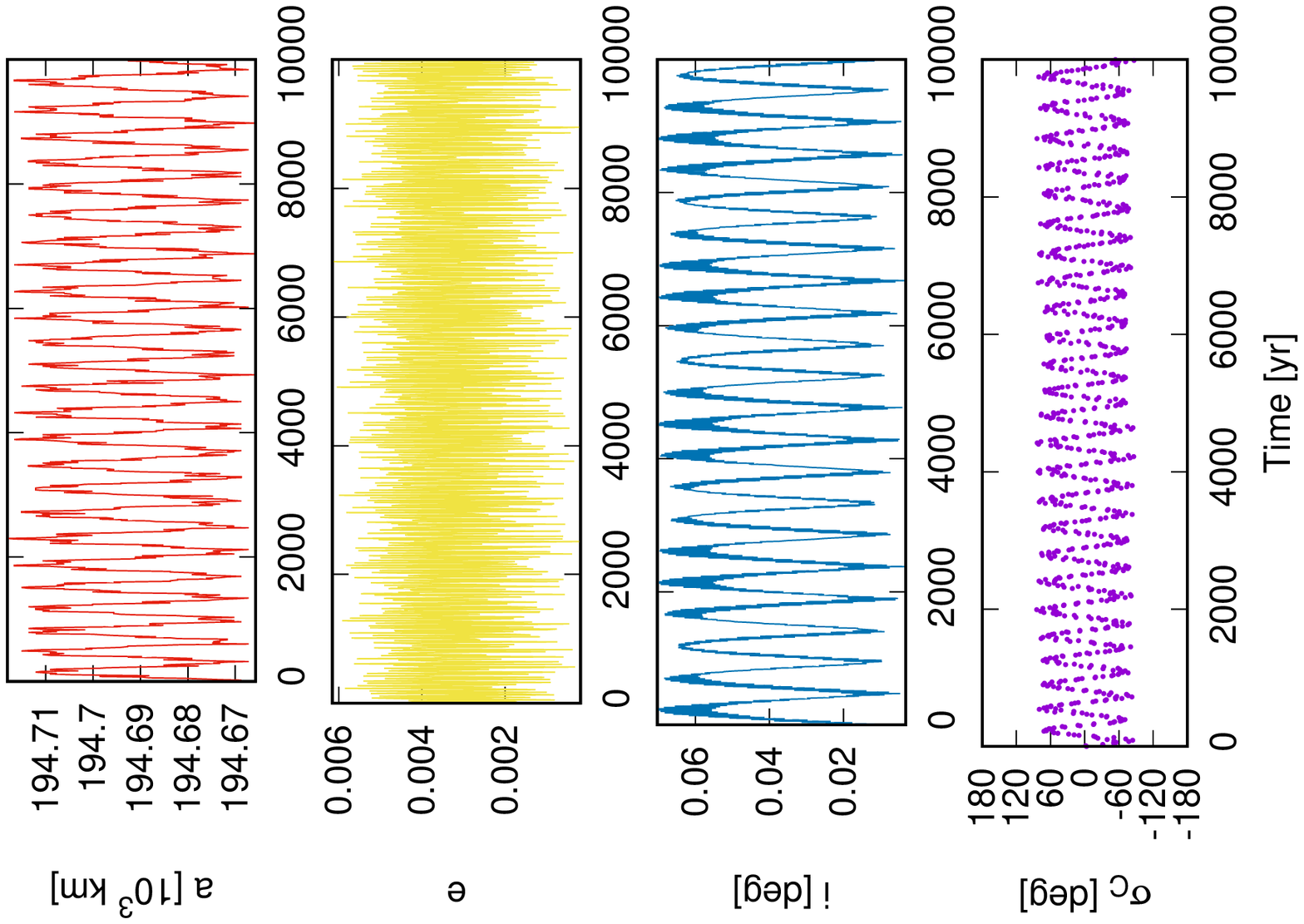}
\includegraphics[width=0.6\columnwidth,angle=270]{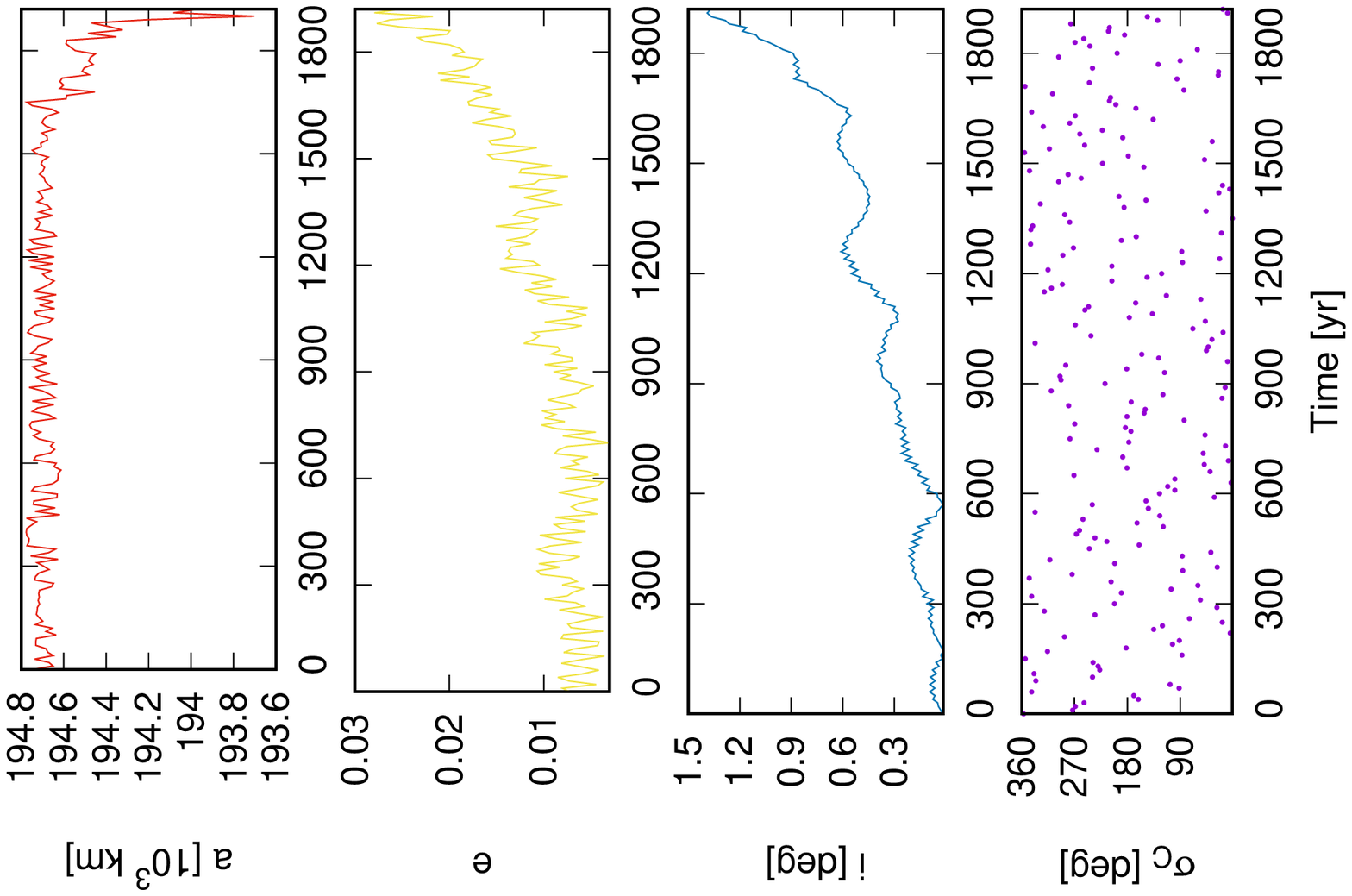}
\caption{ Time variation of semi-major axis, eccentricity, inclination and critical angle $\sigma_C$ for two test particles in the vicinity of Methone. \textit{Left}: the initial condition is $a=194,678$ km and $e=0.006$ (inside of the CER zone). All initial angles are equal to the angles of Methone for the adopted date. This particle is trapped into the 14/15 CER with Mimas, as shown by the libration of $\sigma_C$, surviving for $10^4$ yr. \textit{Right}: The starting values are $a=194,748$ km and $e=0.0073$ (outside of the CER zone). The semi-major axis of the particle decreases 781 km, approaching to Mimas, whereas the eccentricity and inclination increases and $\sigma_C$ circulates. This particle ultimately collides with Mimas in 1923 yr.}
\label{elem-methone}
\end{center}
\end{figure}

On one hand, resonant particles exhibit small excursions of semi-major axis, eccentricity and inclination. On the other hand, outside the limits of the CER zone, the particles attain large variations of orbital elements before collide with Mimas (see Fig. \ref{elem-methone}). None of the simulations resulted in ejection, thus, all instable orbits are represented in this case by collisions of particles with Mimas. Note the large variation of semi-major axis, with values of $\Delta a$ as large as 10,000 km can be obtained. Note also that only a small amount of final orbits, initially outside the limits of the CER, remains stable within the adopted integration time, mainly for $e<0.005$.


Fig. \ref{elem-methone} shows the orbital evolution of two test particles with initial conditions such that one of them stars within the resonant region, whereas the other one stars outside that region. Note as the particle is trapped into the corotation resonance, with a libration of the critical angle $\sigma_C$ around $0^{\circ}$ with an amplitude close to $80^{\circ}$. It is worth emphasize that the orbital evolution within the CER zone depends on the specific location on the $(a,e)$ space of initial conditions, thus, the amplitudes and periods of oscillations of orbital elements and critical angle of different test particles can vary in a significant way. The test particle starting beyond the resonant domain is subject to strong perturbations in a short timescale, ultimately colliding with Mimas at around 1,900 yr (see Fig. \ref{elem-methone} for details of initial conditions).

We thus note as the CER acts as a protection mechanism against collisions for those particles evolving with motions confined to resonant trapping with Mimas. The result of numerical simulations indicates that the 14/15 corotation resonance between Methone and Mimas is dynamically stable for at least $10^4$ yr. It is worth to note that this result is not restricted to Methone itself but is valid for any massless body trapped in the resonant region. 

\subsection{Anthe}
\label{anthe}

\begin{figure}
\begin{center}
\includegraphics[width=0.55\columnwidth,angle=270]{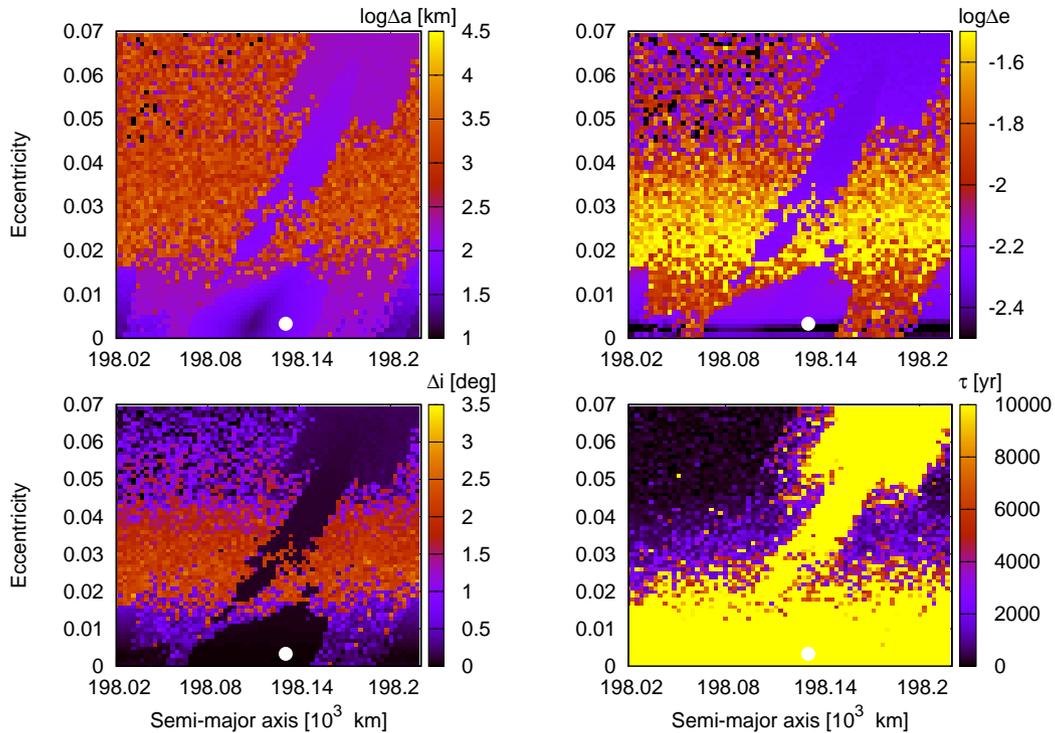}
\caption{ The same than Fig. \ref{mapa-methone} here for Anthe. In this case we note the presence of both corotation (bottom centre) and Lindblad (top right) 10/11 resonances with Mimas. The filled dot indicates the values of Anthe for the adopted date.}
\label{mapa-anthe}
\end{center}
\end{figure}

We repeated here the same analysis of the previous section now for the satellite Anthe. Fig. \ref{mapa-anthe} displays the results, where the initial values
of semi-major axis and eccentricity of test particles range in 198,020 $\le a \le$ 198,220 km and 0 $\le e \le$ 0.07, respectively. The regions of corotation and Lindblad (hereafter, LER) external resonances are clearly seen, corresponding to the 10/11 MMR with Mimas \citep[see][for a detailed description of these resonances]{callegari2020}.
{The Lindblad resonance is characterized by the libration of the critical angle involving the pericentre of the perturbed body, in this case given by $\sigma_L=11\lambda-10\lambda_{Mimas}-\varpi$}. The number of surviving particles for $10^4$ yr is 2316, representing $47.3\%$ of the total sample. Anthe, which is represented by the filled circle, is located inside the corotation zone which extends closely between 198,075 km and 198,150 km and for $e<0.02$. Regarding the stability, we note that almost all particles initially within the CER and LER survive for $10^4$ yr. Despite that there seems to be a connection between the CER and LER regions, these two resonances are separated by a few km \citep{callegari2020}. We note as the eccentricity and orbital inclination suffer the larger variations for initial eccentricities such that $0.02<e<0.04$, except when particles are trapped in the LER. Note also the horizontal structure in the panel for $\Delta e$ in Fig. \ref{mapa-anthe} as in the case of Methone. Anthe is located inside this very stable region and the width of the corotation region also decreases as the eccentricity increases.

{Note that, as shown for the survival time in Fig. \ref{mapa-anthe}, those particles evolving outside of
the limits of the corotation and Lindblad resonances, but close to its border, have unstable motion,
resulting in high increase in their orbital eccentricities. This result has been already discussed in
\cite{callegari2020}, where the domains of the Lindblad resonance in the Anthe's phase space have been first mapped.
In addition, we note from the $\Delta e$ panel that particles evolving in the Lindblad region have larger values of $\Delta e$ than those evolving close to the position of Anthe (white point).}

\begin{figure}
\begin{center}
\includegraphics[width=0.6\columnwidth,angle=270]{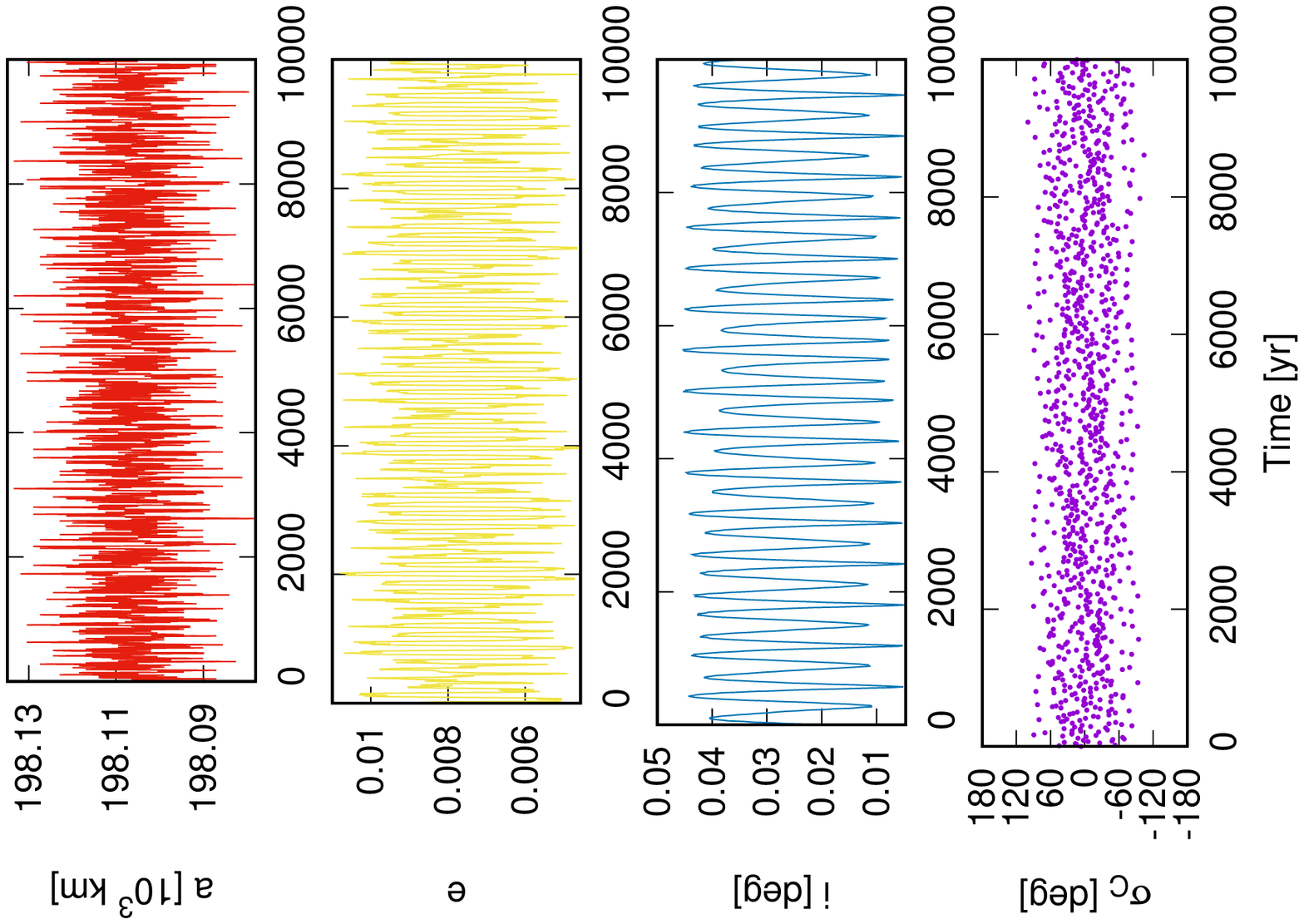}
\includegraphics[width=0.6\columnwidth,angle=270]{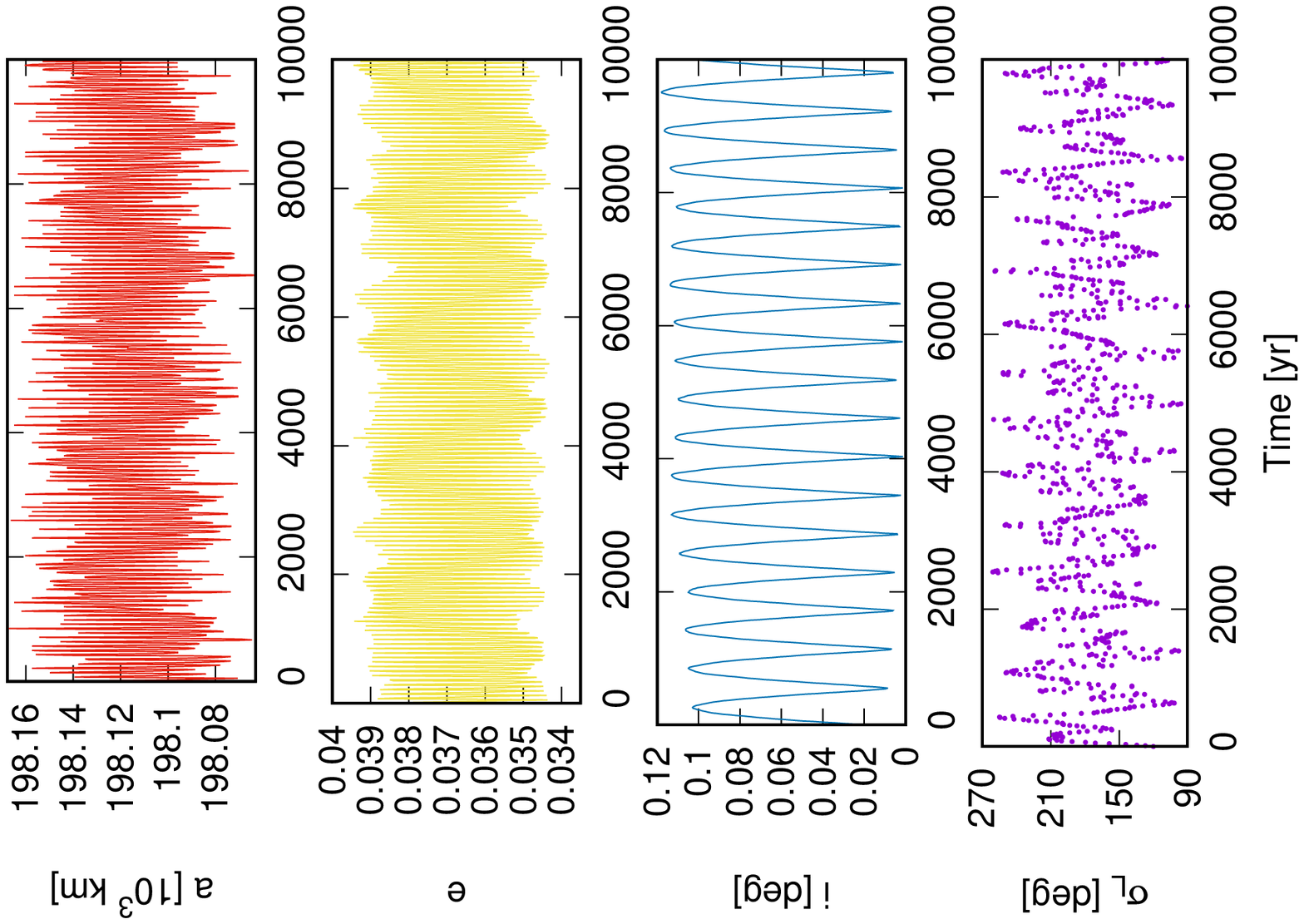}
\caption{ Time variation of semi-major axis, eccentricity, inclination and critical angles for two test particles in the vicinity of Anthe evolving within the CER and LER with Mimas. All initial angles are equal to the angles of Anthe for the adopted date. The particles are trapped into the 10/11 CER (left) and LER (right) with Mimas, as shown by the libration of $\sigma_C$ and $\sigma_L$, surviving for $10^4$ yr. \textit{Left}: the initial condition is $a=198,135$ km and $e=0.01$ (inside of the CER zone). The critical angle librates around $0^{\circ}$. \textit{Right}: The starting values are $a=198,148$ km and $e=0.039$ (inside of the LER zone). The critical angle librates around $180^{\circ}$ \citep[see also][]{callegari2020}.}
\label{elem-anthe}
\end{center}
\end{figure}

Fig. \ref{elem-anthe} shows the time variation of semi-major axis, eccentricity, inclination and critical angle of two test particles evolving
under trapping into the CER (left panel) and LER (right panel) with Mimas. We note the small variations of the orbital elements and the libration of both critical angles $\sigma_C=11\lambda-10\lambda_{Mimas}-\varpi_{Mimas}$ and $\sigma_L=11\lambda-10\lambda_{Mimas}-\varpi$ around $0^{\circ}$ and $180^{\circ}$, respectively. Both particles exhibit very regular motions within the adopted integration time (see Fig. \ref{elem-anthe} for details). For sake of completeness we also checked the behaviour of $\sigma_L$ and $\sigma_C$ for the test particles evolving in the CER and LER regions, respectively. However, both angles circulates with time.

As in the case of Methone, the results of numerical simulations indicates that the 10/11 corotation resonance between Anthe and Mimas is dynamically stable within $10^4$ yr of orbital evolution.

\subsection{Aegaeon}
\label{aegaeon}

\begin{figure}
\begin{center}
\includegraphics[width=0.55\columnwidth,angle=270]{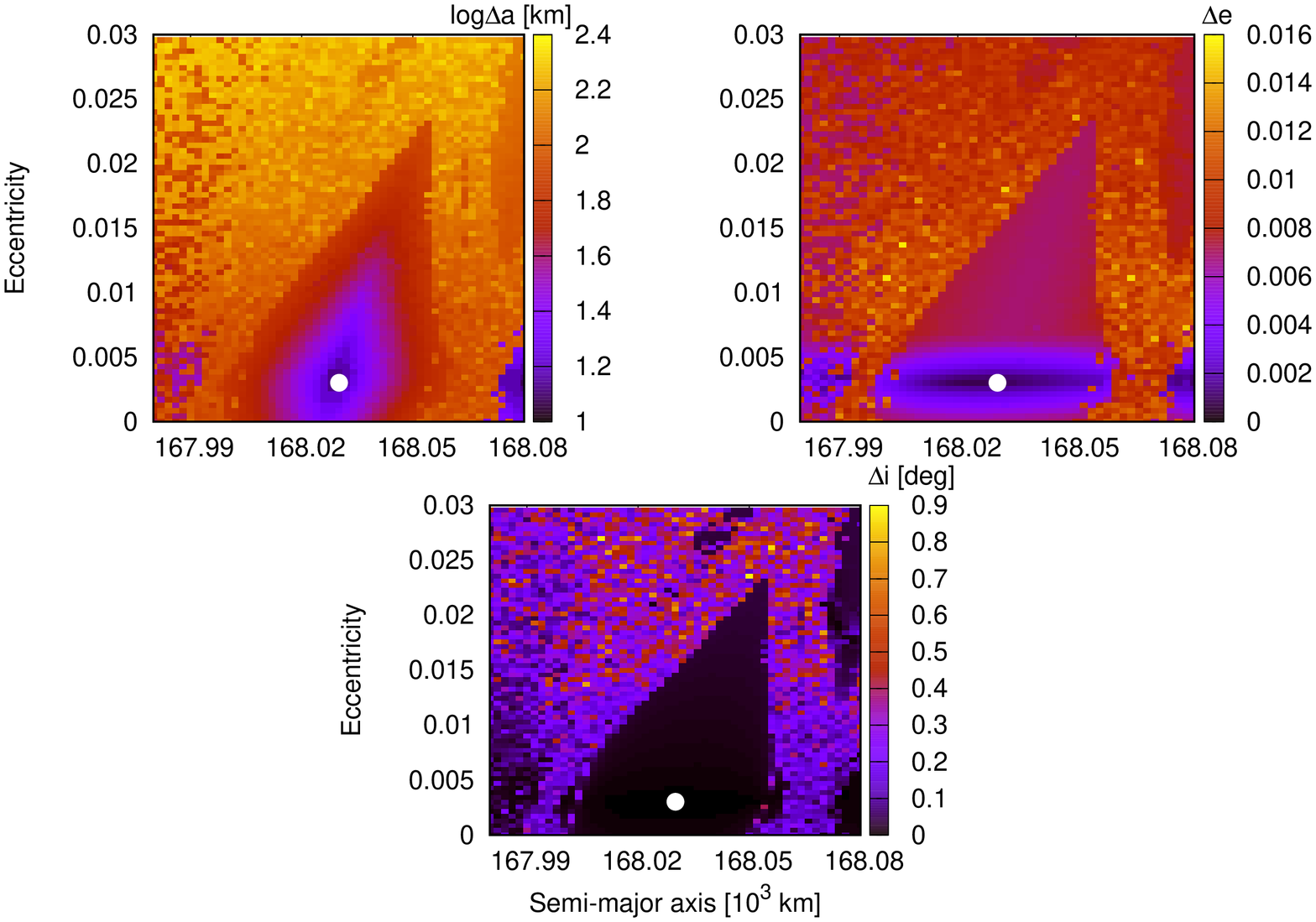}
\caption{ The same than Fig. \ref{mapa-methone} here for Aegaeon. The 7/6 corotation resonance with Mimas appears in the three panels. We do not include the survival time because almost all particles survive for the whole integration time. Note as Aegaeon, identified with the white filled dot, is deeply inserted in the resonance. The results of this figure are in agree with Callegari et al. (2021, in preparation).}
\label{mapa-aegaeon}
\end{center}
\end{figure}

Fig. \ref{mapa-aegaeon} shows the result of 4,900 test particles in the vicinity of Aegaeon. The lateral bar indicates the values of $\log\Delta a$, $\Delta e$, and $\Delta i$. The initial values of semi-major axis and eccentricity range in 167,980 $\le a \le$ 168,080 km and 0 $\le e \le$ 0.03. The filled circle shows the values of Aegaeon for the adopted initial date. It is important to note that only 3 test particles particles did not remain in stable orbit for $10^4$ yr of orbital evolution, thus, we do not show here the result for $\tau$, the survival time. We clearly see the corotation zone associated to the internal 7/6 CER with Mimas, closely between 168,010 km and 168,060 km and $e<0.021$. Aegaeon (white dot) is deeply inserted in the resonance, which has its width reduced for high values of the eccentricity.

\begin{figure}
\begin{center}
\includegraphics[width=0.6\columnwidth,angle=270]{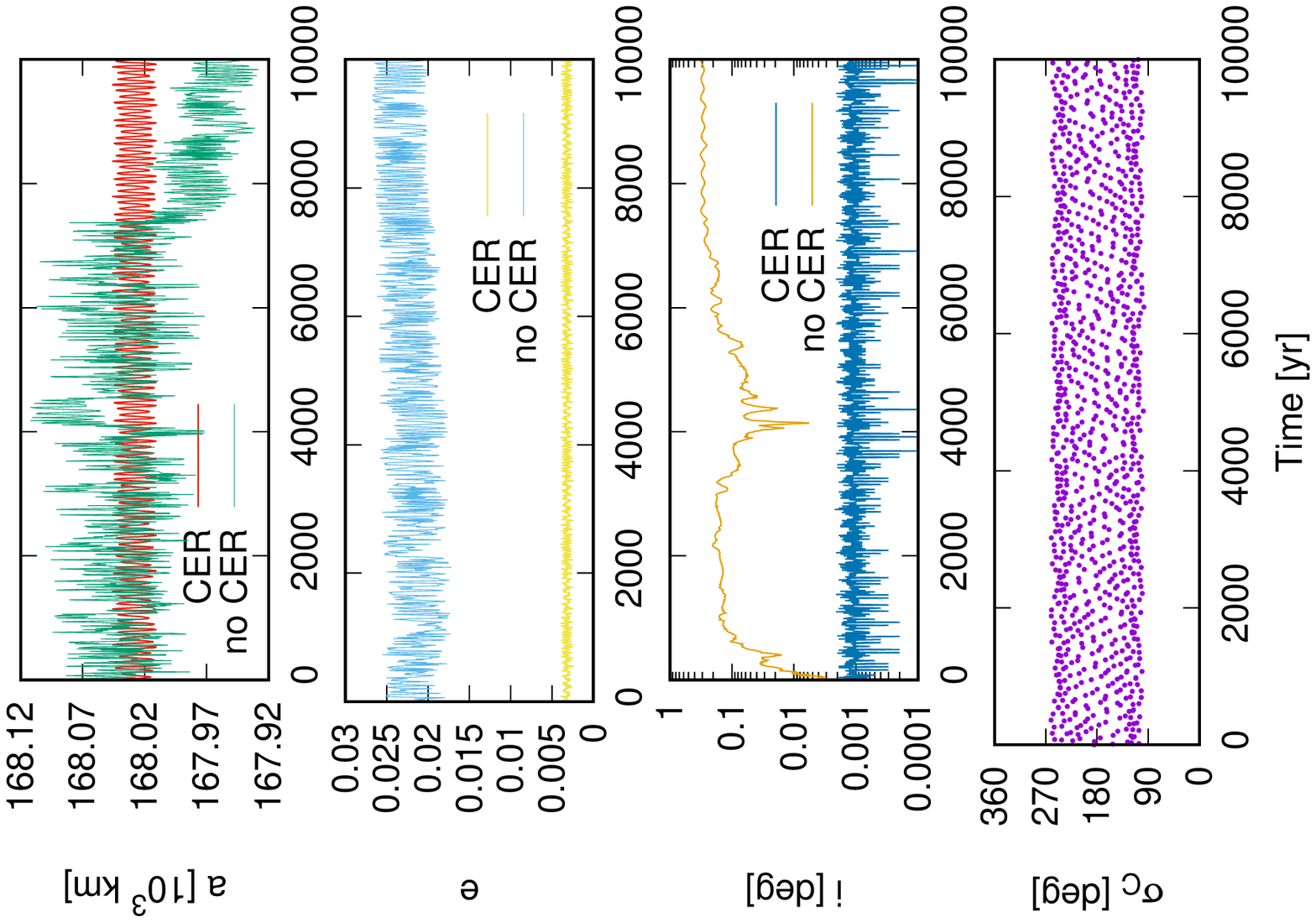}
\caption{ Time variation of semi-major axis, eccentricity, inclination and critical angles for two test particles in the vicinity of Aegaeon evolving under resonant and non-resonant motion associated to the 7/6 CER with Mimas. The particle starting within the domain of the resonance have $a=168,043$ km and $e=0.003$, whereas the particle outside this domain starts with $a=168,043$ km and $e=0.0253$. The critical angle librates around $180^{\circ}$ with amplitude closely to $90^{\circ}$ (the critical angle of the non resonant particle circulates and thus is not shown here).}
\label{elem-aegaeon}
\end{center}
\end{figure}

It is worth emphasize that, differently from Methone and Anthe where surviving particles are restricted to the resonant regions, in the case of Aegaeon almost all particles survive for $10^4$ yr for the adopted grid of initial conditions. This fact indicates that the neighborhood of Aegaeon is more dynamically stable than for the other satellites also trapped in resonant motion with Mimas (the situation is different when the mass of Aegaeon is taken into account, as it is described in Sec. \ref{massive}). However, as can be seen in Fig. \ref{elem-aegaeon}, numerical simulations starting outside the CER result in significant variations of orbital elements. 

We also note, in the top right panel of Fig. \ref{mapa-aegaeon}, the horizontal bar structure on the bottom part of the corotation zone. Moreover, Aegaeon is located at the middle of this structure, which is also associated to the maximum width of the resonance.


Fig. \ref{elem-aegaeon} shows the time variation of the orbital elements and the critical angle of the 7/6 CER resonance with Mimas, namely, $\sigma_C=7\lambda_{Mimas}-6\lambda-\varpi_{Mimas}$, for two evolving particles. One of these particles starts within the domain of the CER and the other one stars 
outside the resonance. Note as the particle evolving out of resonance have large excursions of semi-major axis, eccentricity and inclination when compared with the resonant one. We also note that the particle evolving in the CER shows the libration of $\sigma_C$ around $180^{\circ}$ with an amplitude of $\simeq90^{\circ}$. 
The numerical simulations show that the 7/6 corotation resonance between Aegaeon and Mimas have a dynamical stability time larger than $10^4$ yr.

\subsection{Pallene}
\label{pallene}

\begin{figure}
\begin{center}
\includegraphics[width=0.5\columnwidth,angle=270]{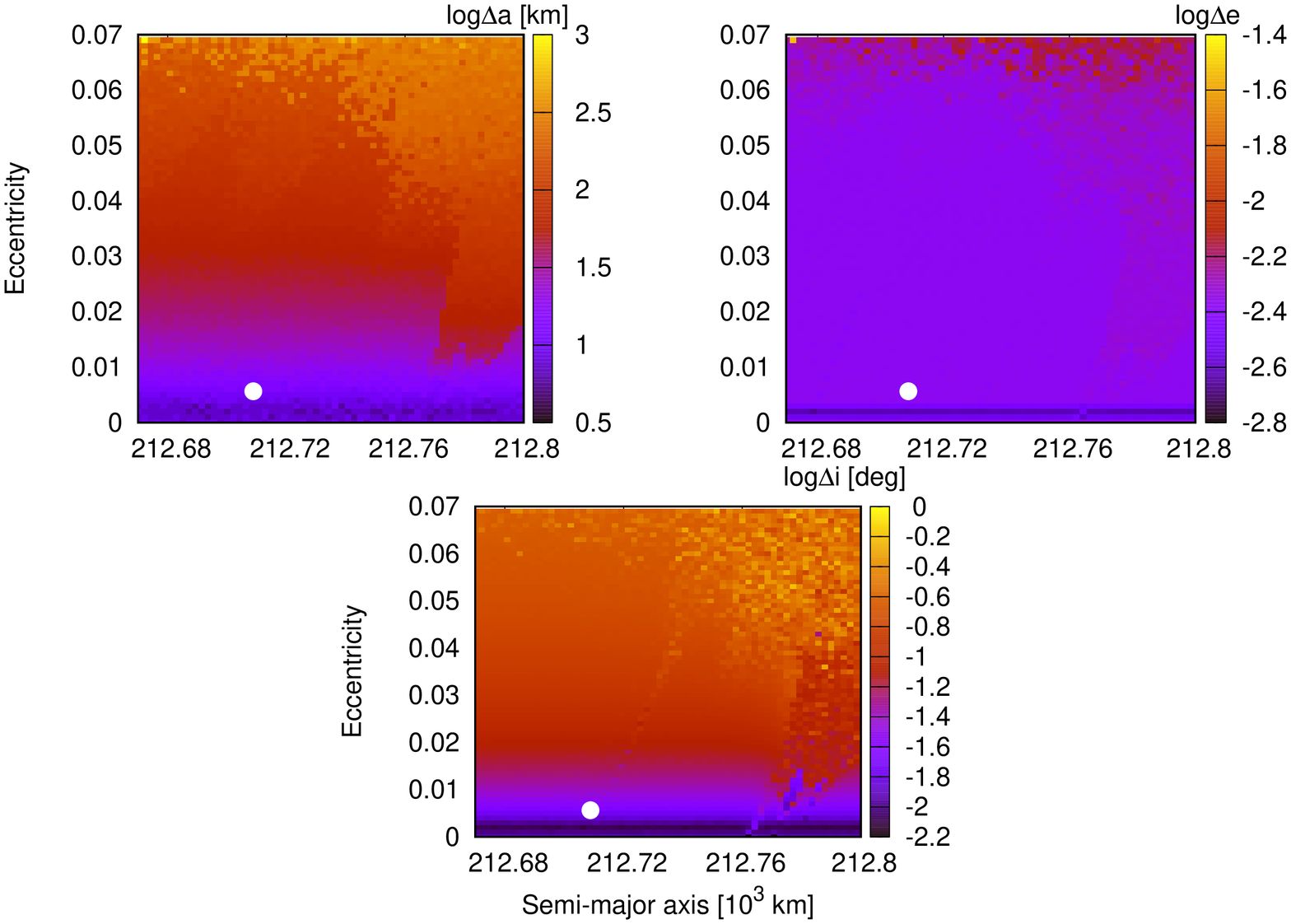}
\caption{ The same than Fig. \ref{mapa-methone} here for Pallene. As in the case of Aegaeon, we do not show the survival time because all the test particles evolve without instabilities. No significant structure is observed in this case, as Pallene is not currently evolving under resonant capture with a companion satellite. The white filled dot indicates the values of Pallene for the adopted date.}
\label{mapa-pallene}
\end{center}
\end{figure}

Despite that \citet{spitale+2006} suggested that Pallene could be trapped in the 19/16 MMR with Enceladus, this satellite is not currently evolving in any 
resonant motion. The resulting orbital evolution for a set of test particles in the vicinity of Pallene is shown in Fig. \ref{mapa-pallene}, where initial semi-major axis and eccentricity range in $212,670\leq a\leq212,820$ km and $0\leq e\leq0.07$. Differently from previous satellites, we do not observe any structure such as regions identified with orbital resonances. The motion of all the 4,900 test particles resulted in stable orbits along $10^4$ yr of orbital evolution. The results for $\Delta a$, $\Delta e$ and $\Delta i$ indicates that the orbital elements have small variations for $e\leq0.01$. Pallene, identified with the filled circle, is currently located in this region of small variations of the orbital elements, as can be seen in Fig. \ref{mapa-pallene}.


\subsection{Remarks}

It is important to mention that there is possible to make theoretical predictions about the widths of the corotation resonances 
\citep[e.g.][]{elmutamid+2014,AHearn+2019}. Thus, we compute the widths of the three CERs by using the equation (7) of \cite{AHearn+2019}, obtaining $\simeq$ 112 km, 97 km and 59 km for Methone, Anthe and Aegaeon, respectively. These results are in very good agreement with the radial sizes of the corotation zones obtained from the numerical simulations, as can be seen in Figs. \ref{mapa-methone}, \ref{mapa-anthe}, \ref{mapa-aegaeon}. Note that these results also agrees with the ones reported in \cite{elmutamid+2014}\footnote{The widths obtained in \cite{elmutamid+2014} are the half-widths.} and \cite{madeira+winter2020}.

We have defined the survival time as the one for which the test particles becomes in unstable motion. However, for those particles surviving the entire simulation, it is not appropriate to define the survival time as the simulation length, which should be considered as a lower limit of the survival time. For instance, in Figs. \ref{mapa-methone} and \ref{mapa-anthe}, the time $10^4$ yr represents this lower limit.

Note that some particles that do not survive can have very small variations of semi-major axis, eccentricity and inclination. However, this is because these particles did not stay in the simulation long enough to have those values changed, thus, introducing some source of bias in the results.

We still mention that, in Figs. \ref{mapa-methone}, \ref{mapa-anthe}, \ref{mapa-aegaeon} and \ref{mapa-pallene} we decided to show only the relevant ranges of initial conditions for each small moon because the widths and vertical extent (on the space $a$\,-\,$e$) of the corotation resonances depends on the specific satellite. The reason is purely for convenience in order to show a better representation.

\subsection{Clumps}
\label{clumps}

\begin{figure}
\begin{center}
\includegraphics[width=0.33\columnwidth,angle=270]{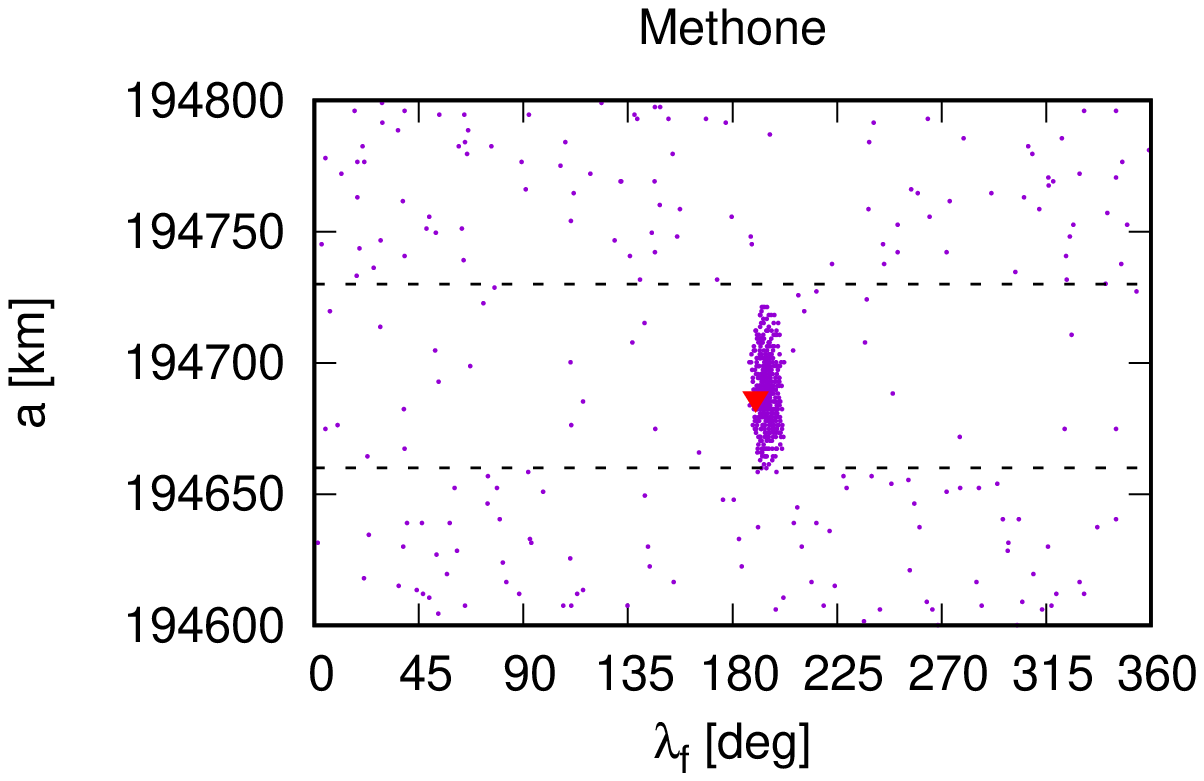}
\includegraphics[width=0.33\columnwidth,angle=270]{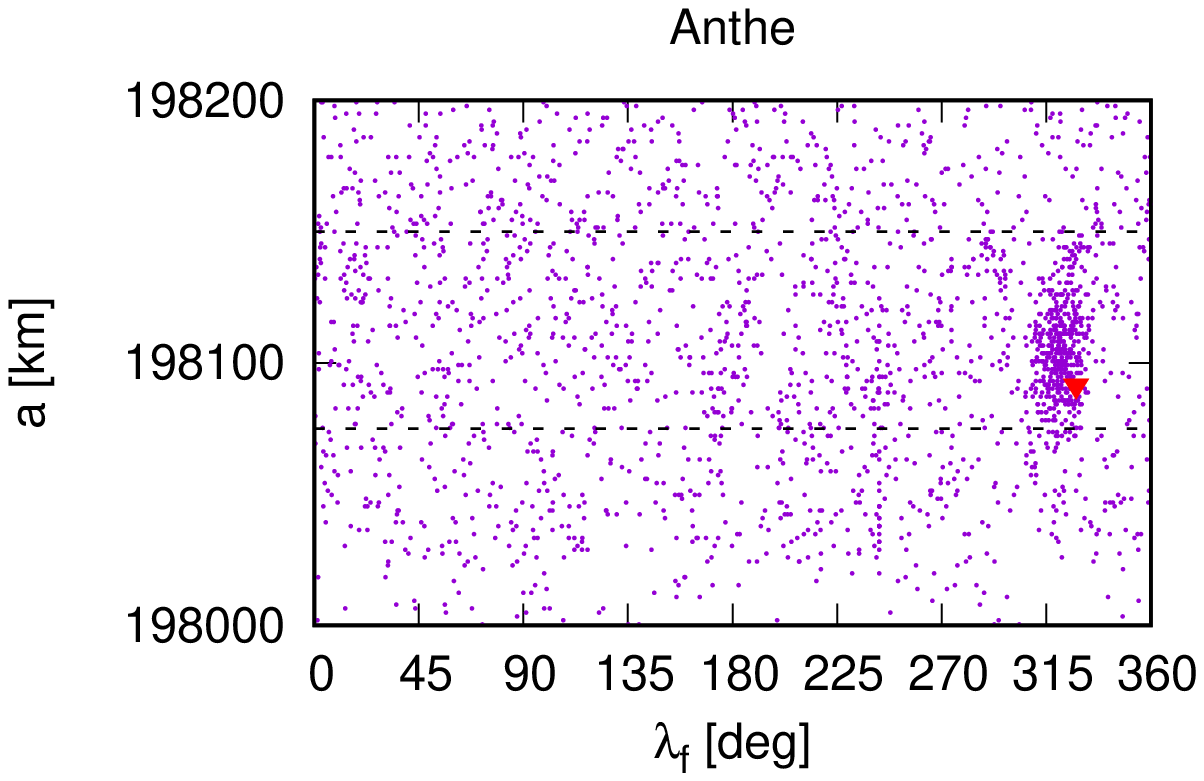}
\includegraphics[width=0.33\columnwidth,angle=270]{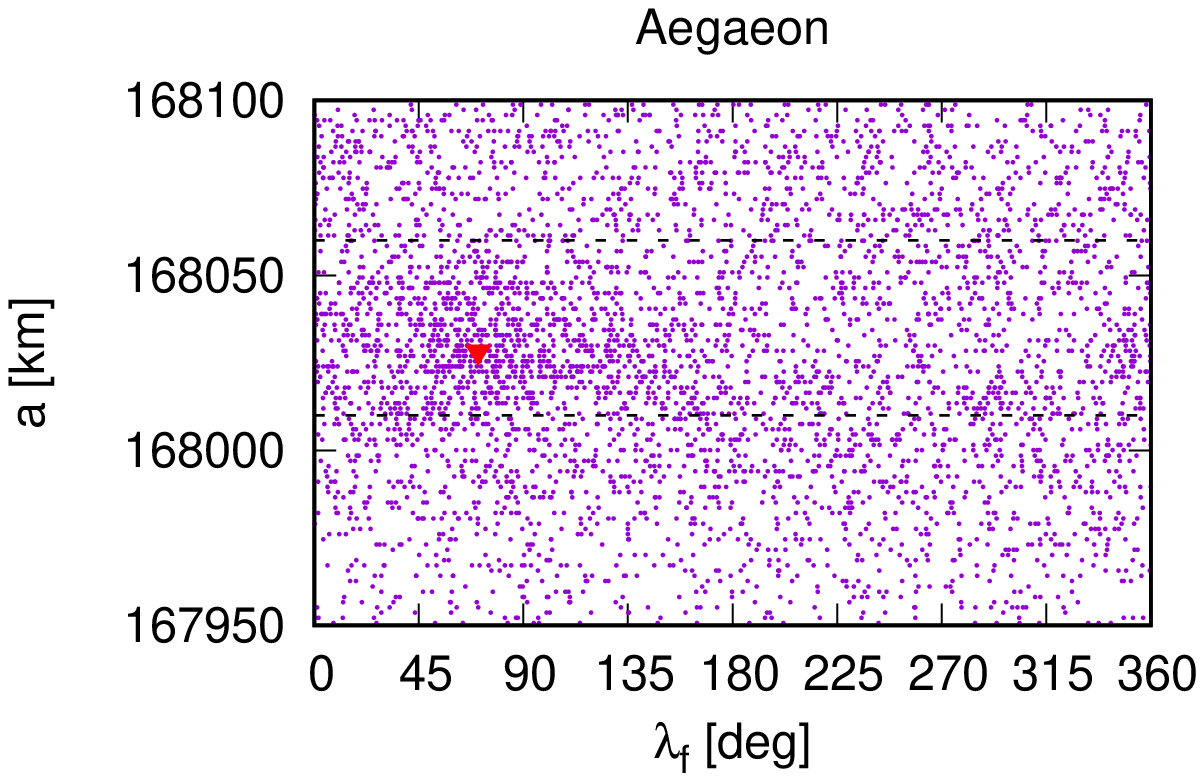}
\includegraphics[width=0.33\columnwidth,angle=270]{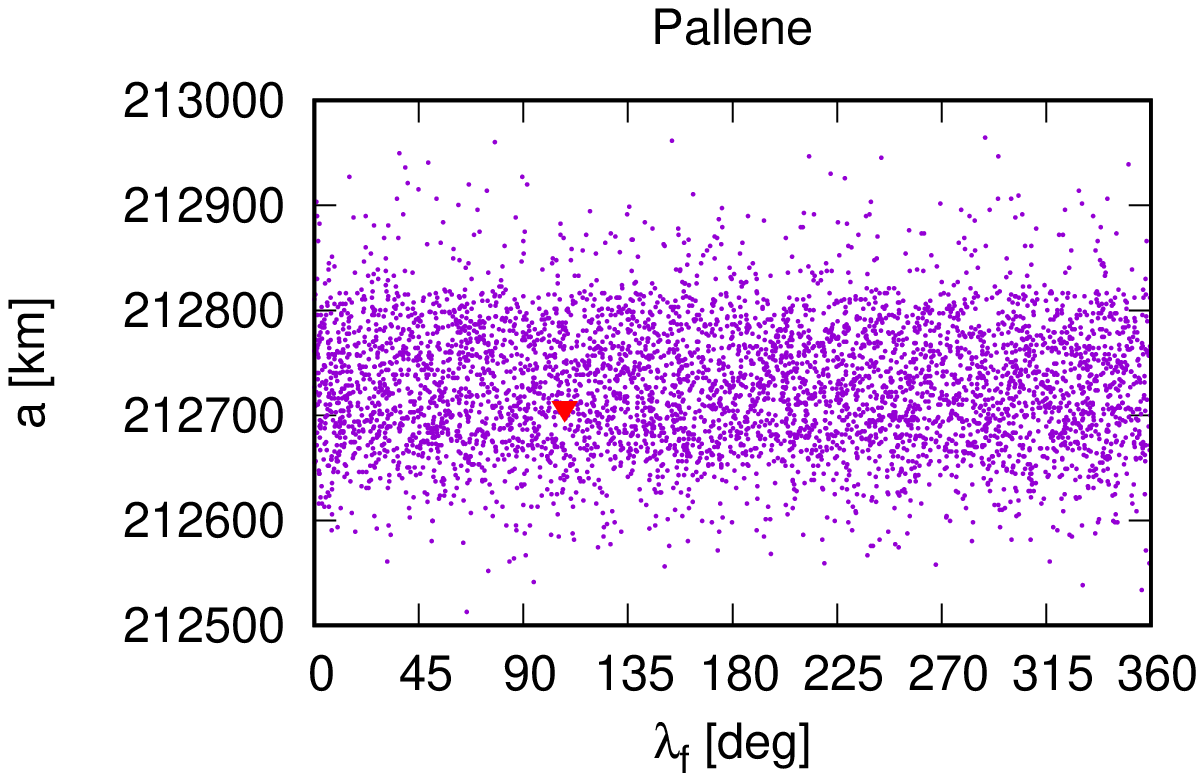}
\caption{Final distribution of semi-major axis and mean longitudes of massless particles in the vicinity of the corotation resonances with Mimas for Methone, Anthe and Aegaeon in the case of massless satellites. For sake of completeness we also show the final distribution for the non resonant satellite Pallene. {The dashed black lines indicate the locations and widths of the corresponding corotation resonances, according to Figs. \ref{mapa-methone}, \ref{mapa-anthe} and \ref{mapa-aegaeon}}. Clumps of particles are identified for the resonant small moons with longitudinal extents similar to the observed arcs of Methone and Anthe. In the case of Aegaeon, the clump is less visible because the dynamical environment is more stable, however, the width of the clump have a similar size to the observed arc in the Saturn G ring (see text for details). Red triangles indicate the final values of semi-major axis and mean longitudes of Aegaeon, Methone, Anthe and Pallene.}
\label{arcos1}
\end{center}
\end{figure}

Fig. \ref{arcos1} shows, for each small moon, the final semi-major axis as a function of the final mean longitude of all particles on the grid of numerical simulations. In each panel, the dashed black lines indicate the locations and widths of the corresponding corotation resonances, whereas the red triangles indicate the final values for the small moons. It is worth note the clumps of particles for Methone, Anthe and Aegaeon. These clumps are restricted to the domain of the respective CER and have different widths in longitude. For Methone, the clump width is $\Delta\lambda\sim12^{\circ}$; for Anthe is $\Delta\lambda\sim25^{\circ}$ and for Aegaeon is $\Delta\lambda\sim60^{\circ}$. The particles of the clumps survives until the end of the simulation ($10^4$ yr), as they belong to the CER zones. 

Consider a particle $i$ trapped in a $(p+1)/p$ CER with Mimas with a critical angle $\phi_i=(p+1)\lambda_i-p\lambda_{Mimas}-\varpi_{Mimas}$ (for $p>1$ and integer). Here $\lambda_{Mimas}$ and $\varpi_{Mimas}$ are the mean and pericentre longitude of Mimas, respectively. We have, for the $i$ particle

\begin{equation}
\lambda_i=\frac{1}{p+1}(\phi_i+p\lambda_{Mimas}+\varpi_{Mimas}). 
\end{equation}
Applying the same reasoning for an additional particle $j$ also trapped in the same CER, we finally have

\begin{equation}\label{eq-arcos}
\lambda_j-\lambda_i=\frac{1}{p+1}(\phi_j-\phi_i).
\end{equation}
Eq. (\ref{eq-arcos}) shows that, since $\phi_i$ and $\phi_j$ oscillate because both particles are trapped in the same resonance, the difference of mean longitudes of any pair of particles trapped in the CER also oscillates around some value. Hence, test particles evolving within the CER have mutually confined values of their mean longitudes, confirming our numerical results concerning the clumps. 

As we mention in the Sec. \ref{intro}, Methone, Anthe and Aegaeon have arcs structures centred at the small moons. According to \citet{hedman+2009}, the longitudinal sizes of the arcs are $\sim10^{\circ}$ and $\sim20^{\circ}$ for Methone and Anthe, respectively, whereas the arc of Aegaeon extends over $\sim60^{\circ}$ in longitude \citep{hedman+2007}. By mean of analytical predictions, \citet{hedman+2009} also pointed out that the corotation resonances act as a mechanism to confine the particles of the arcs in longitudinal extent. Here we are confirming these features through a numerical exploration regarding the orbital stability of test particles in the vicinity of the different corotation resonances with Mimas. However, we are not establishing a direct relationship between the clumps obtained in our simulations with the observed arcs. Indeed, previous works have investigated the formation and sustainability of arcs structures around Aegaeon, Methone and Anthe considering gravitational and non-gravitational perturbations \citep{araujo+2016,sun+2017,madeira+2018}. In Sec. \ref{arcs} we address this issue in more details.

It is important to note that, in the numerical simulations of this section, we are considering that Aegaeon, Anthe, Methone and Pallene are massless bodies, thus, neglecting their gravitational perturbation with the rest of the particles. The consideration of massive small moons can significantly modify the pattern of final distribution of semi-major axis and mean longitudes of the surviving particles. We will consider this aspect in the next section.

{We remark that not all the plots in Fig. \ref{arcos1} are strictly comparable because different ranges of eccentricities have been used in each case.}

\section{Numerical simulations with massive satellites}
\label{massive}

\begin{table}
\begin{center}
\caption{Mean densities and derived masses for Aegaeon, Methone, Anthe and Pallene. The mean densities are taken from \citet{thomas+2013}, except for Anthe where we adopt the value of \citet{munhoz+2017}. We also refer to \citet{thomas+2020} for properties of the small inner satellites of Saturn using final \textit{Cassini} image data.
The masses are computed from $\rho=3m/4\pi a'b'c'$, where $\rho$ and $m$ are the mean density and mass of the corresponding satellite, respectively, whereas $a'$, $b'$ and $c'$ are the semi-axes of the ellipsoidal shape of each small satellite, which are also taken from \citet{thomas+2013}. In the case of Anthe, only the mean radius is constrained by the observations \citep{thomas+2013}.}
\vspace*{0.3cm}
\begin{tabular}{c c c}
\hline
Satellite & Density (kg/m$^{3}$) & Mass (kg) \\ 
\hline
Aegaeon & 540 & 7.9$\times10^{10}$ \\
Methone & 310 & 3.9$\times10^{12}$ \\
Anthe & 350 & 1.8$\times10^{11}$ \\
Pallene & 250 & 1.2$\times10^{13}$\\
\hline
\label{tabela-massas}
\end{tabular}
\end{center}
\end{table}

\begin{figure}
\begin{center}
\includegraphics[width=0.33\columnwidth,angle=270]{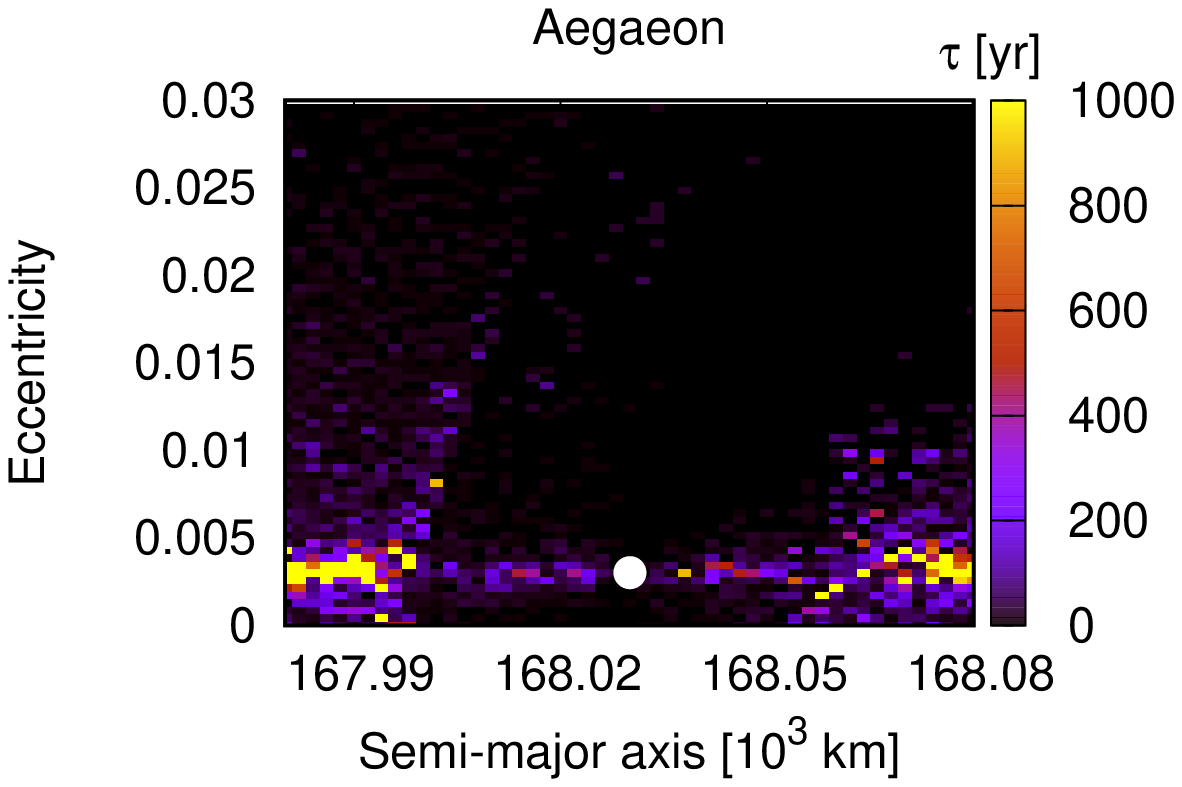}
\includegraphics[width=0.33\columnwidth,angle=270]{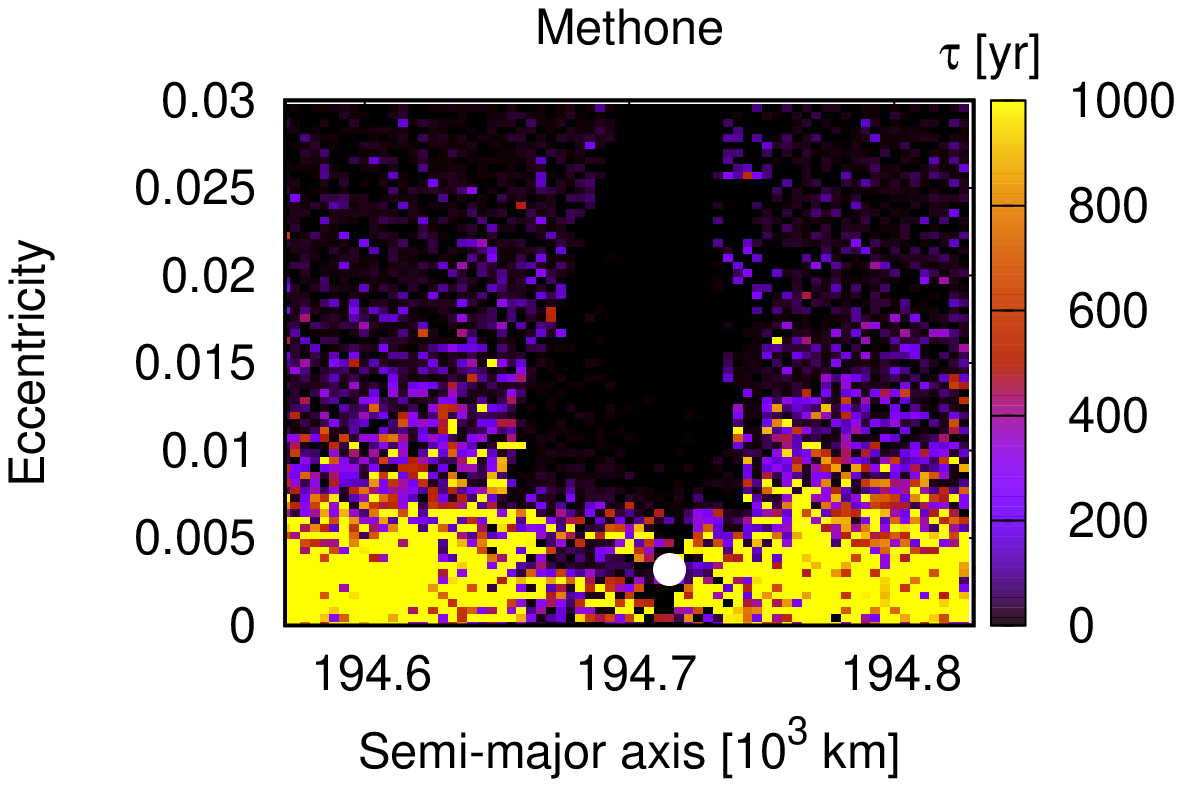}
\includegraphics[width=0.33\columnwidth,angle=270]{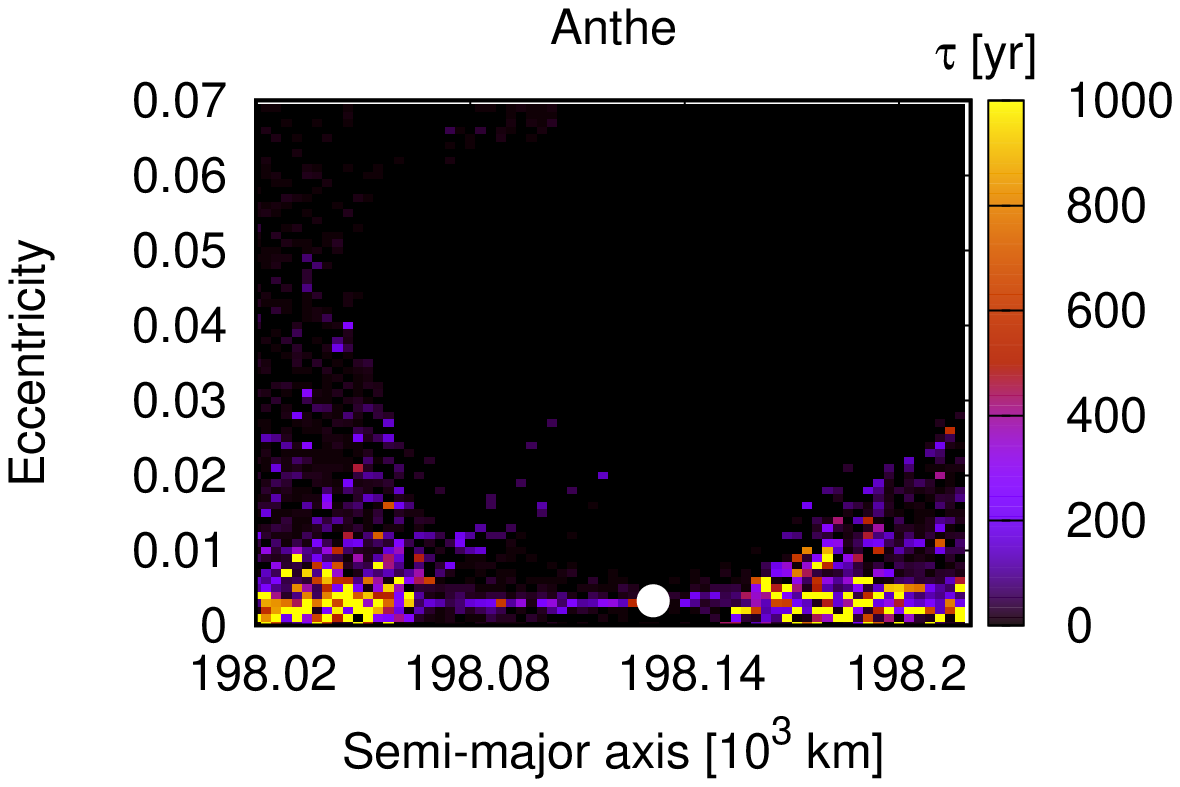}
\includegraphics[width=0.33\columnwidth,angle=270]{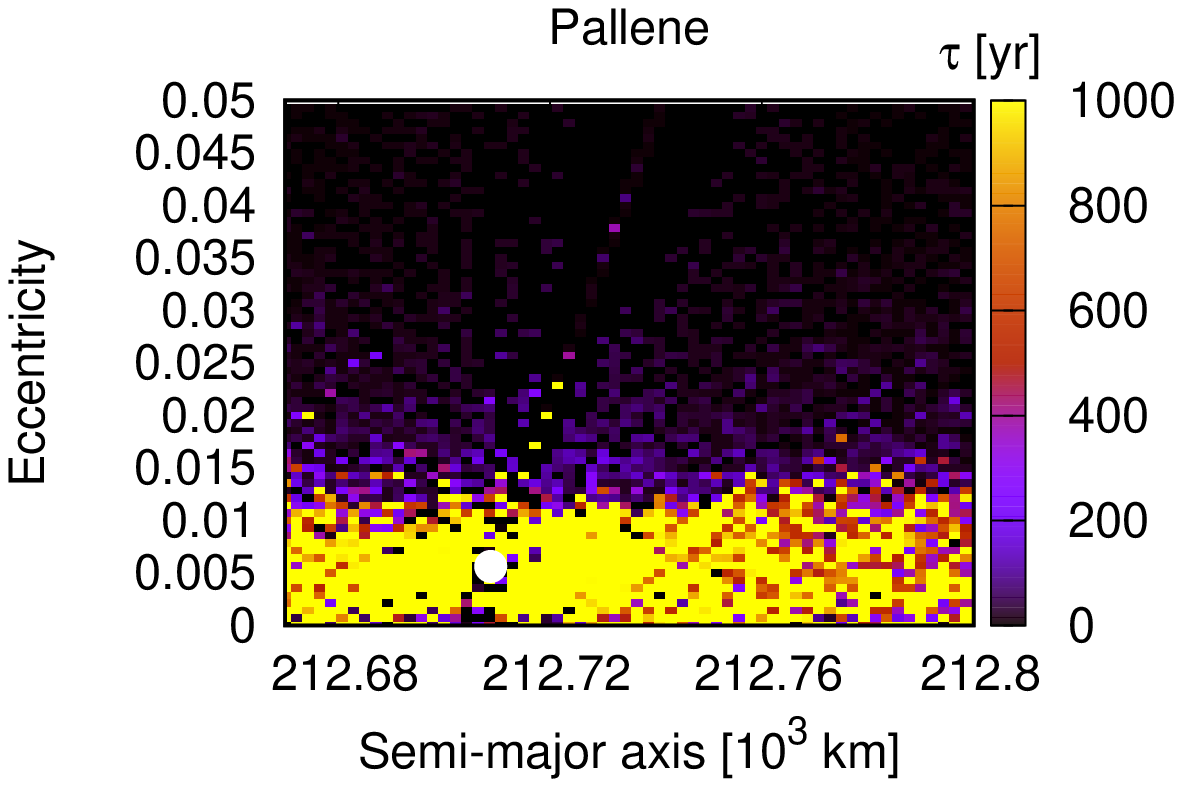}
\caption{ The survival time $\tau$ (in units of years) for four numerical simulations considering 4,900 test particles initially distributed in the vicinity of  Aegaeon, Methone, Anthe and Pallene. Each small moon is considered as a massive body. The initial angles of all test particles are the same that the initial angles of the respective small moon in each simulation. The integration time is $10^3$ yr and the masses are derived from \citet{thomas+2013} (see Table \ref{tabela-massas}). The filled circles represent the values of $a$ and $e$ of each small moon for the adopted epoch, that is, 2020/03/25.}
\label{mapas-massa}
\end{center}
\end{figure}

In this section we consider that Aegaeon, Methone, Anthe and Pallene are massive bodies. However, due to their small masses, we assume that the mutual perturbation between the small moons can be neglected. This allow us to analyze the dynamical behaviour around each small moon in a separate way. We take the values of radii and mean densities from \citet{thomas+2013} and calculate the mass of each satellite, which are shown in Table \ref{tabela-massas}. In the same way that above section, we analyze the dynamical evolution of test particles in the vicinity of the four small moons. Here again the initial angles of the test particles are the same that the angles of the corresponding small moon, however, in Sec. \ref{massive-random} we also consider random initial angles. The results are summarized in Fig. \ref{mapas-massa}, where we show the survival time for 4,900 test particles initially distributed, in the same intervals of semi-major axis and eccentricity that in previous section, in the vicinity of Aegaeon, Methone, Anthe and Pallene. 

Fig. \ref{mapas-massa} shows that the perturbation of the small moons have a severe effect on the motion of surrounding particles, destabilizing most of them in a timescale of a few hundreds of years or even less. In all cases, the instabilities occur due to collisions between the particles with the small moons. Note as the regions associated to corotation resonances are almost depleted of particles (compare with Figs. \ref{mapa-methone}, \ref{mapa-anthe} and \ref{mapa-aegaeon} for $\tau$). We also note that, despite their small masses (see Table \ref{tabela-massas}), Aegaeon and Anthe are more efficient to sweep the test particles from their respective neighborhoods, even for very small initial eccentricity. In fact, for Pallene, particles with initial eccentricities such that $e<0.01$ survive for $10^3$ yr. We remember that Pallene is not currently evolving under resonant trapping with a companion satellite. For Methone, those particles with $e<0.005$ remain in stable orbit for $10^3$ yr, however, they are mostly restricted to the region outside the 14/15 corotation resonance (as in the case of massless Methone, see Fig. \ref{mapa-methone}).

The results above indicate that the orbital dynamics of bodies with negligible mass within the corotation resonances of Aegaeon, Methone and Anthe with Mimas is strongly perturbed by the gravitational interaction with the small moons. 

{As shown in Fig. \ref{mapas-massa} and Table \ref{tabela-survivor}, the percentage of surviving particles is larger for Pallene. This fact indicates that, since Pallene is not in resonance with Mimas, the corotation resonance plays a fundamental role in the dynamical stability of surrounding particles.}

It is worth nothing that \citet{munhoz+2017} also performed a dynamical analysis of the region around Mimas with the aim to explore the influence 
of massive Aegaeon, Methone, Anthe and Pallene on nearby particles, with the implications for the formation of arcs/rings. They cover a large region in semi-major axis and do not focus in the dynamical stability within the CER regions. Moreover, \citet{munhoz+2017} analyze the dynamical environment of the four moons by means of the diffusion map technique and their numerical simulations cover 18 yr of evolution.

We also mention that, for sake of completeness, we performed additional sets of numerical simulations considering the upper and lower limits of the small moons mean densities reported in \cite{thomas+2013}. However, the results does not shown significant differences with the ones obtained for the nominal values of the mean densities.

\subsection{Clumps}
\label{clumps-massa}

As in previous section, we investigate the occurrence of clumps of particles for those satellites currently evolving in CER. However, we have not identified any structure in the final distribution of mean longitudes and semi-major axis, even for those particles that do not survive the whole integration time. Hence, the mass of the resonant satellites would have a significant effect on the formation and sustainability of their corresponding arcs of particles (see discussion of Sec. \ref{arcs}).  


\subsection{Random angles}
\label{massive-random}

As we have noted in the previous sections, when the initial angles of the particles equals the initial angles of the small moon (hereafter we call non-random angles), all particles start inside or near the respective corotation resonance with Mimas. In this section we consider that the initial argument of pericentre ($\omega$) and mean anomaly ($M$) of all test particles are randomly varying between 0 and $360^{\circ}$. We also assume that the orbital inclination ($i$) and longitude of the ascending node ($\Omega$) are initially equal to zero. The objective is to initialize the set of test particles in the vicinity of each small moon but not necessarily close to or in the respective corotation resonance. We will then compare with the results summarized in Fig. \ref{mapas-massa}.

\begin{figure}
\begin{center}
\includegraphics[width=0.25\columnwidth,angle=270]{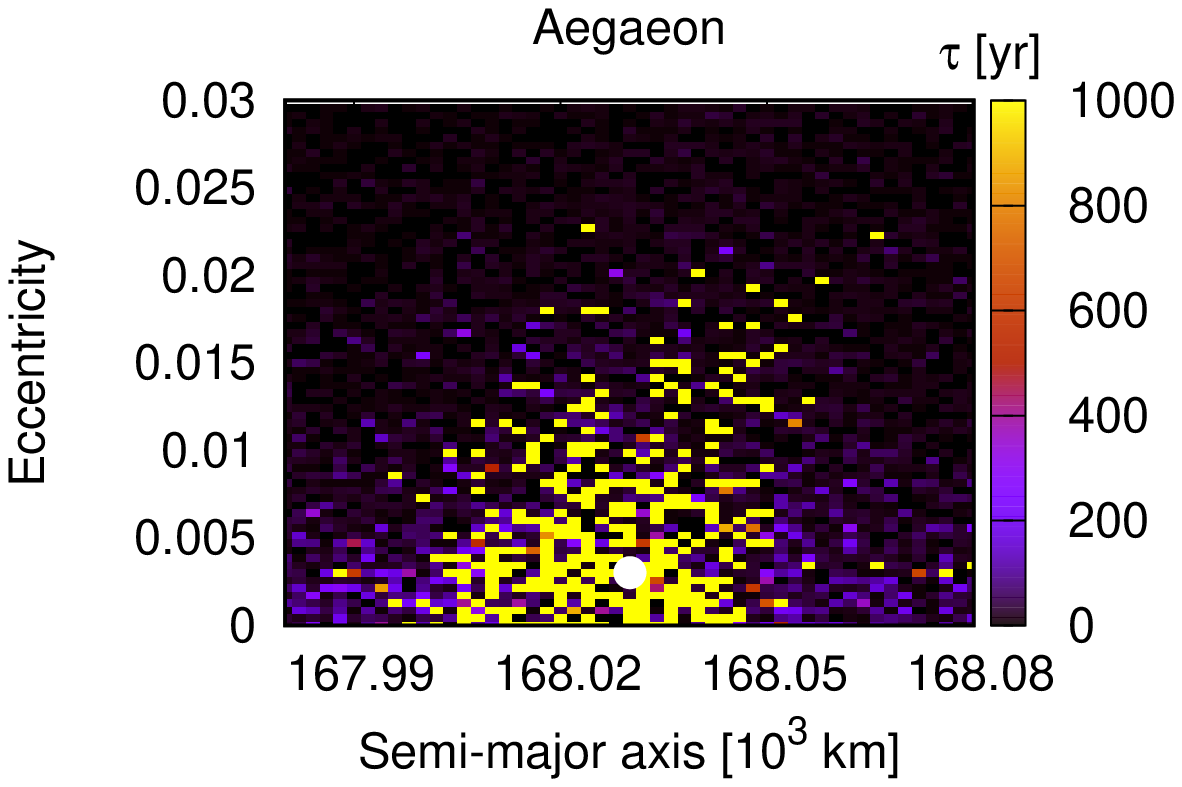}
\includegraphics[width=0.25\columnwidth,angle=270]{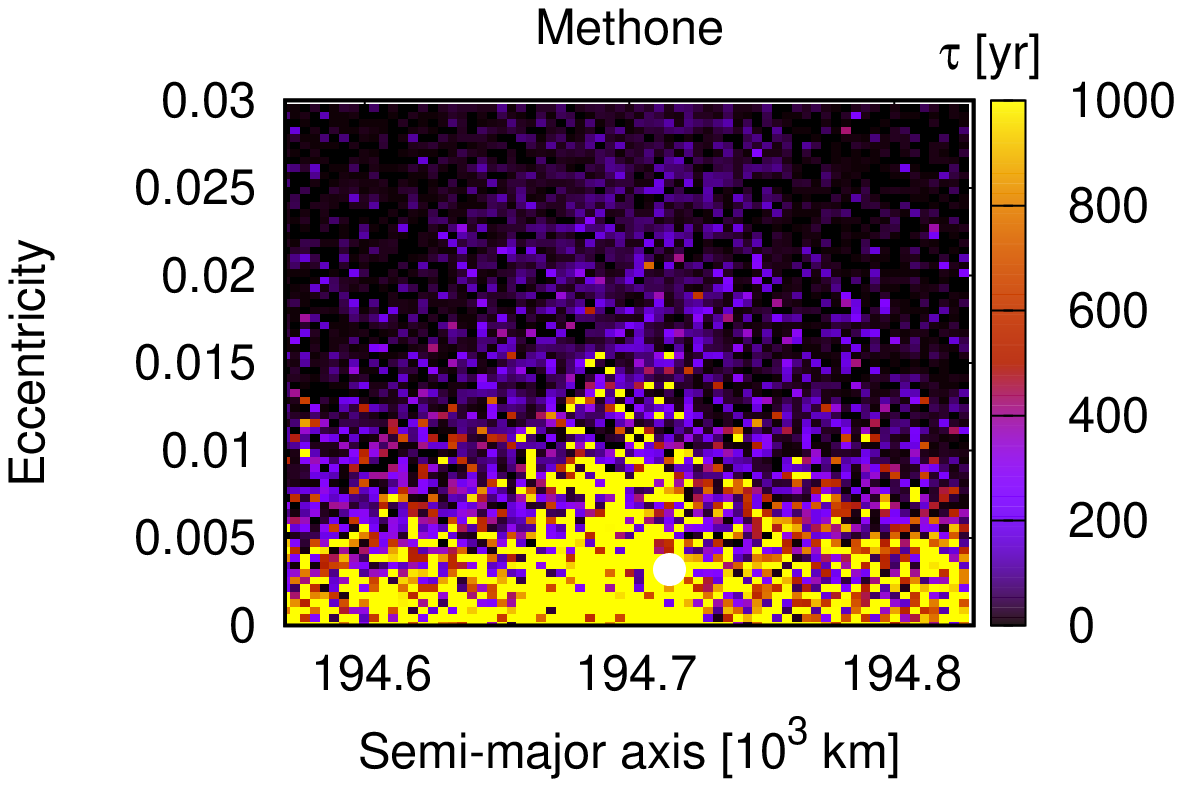}
\includegraphics[width=0.25\columnwidth,angle=270]{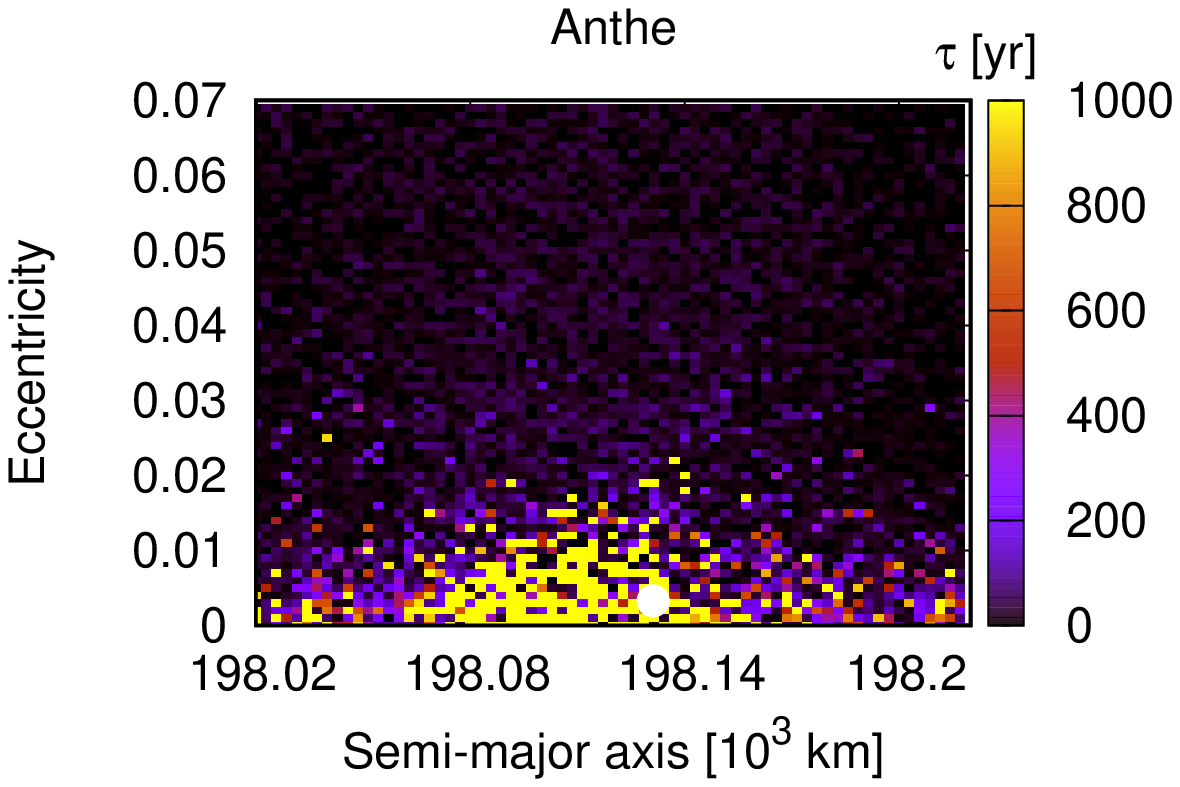}
\includegraphics[width=0.25\columnwidth,angle=270]{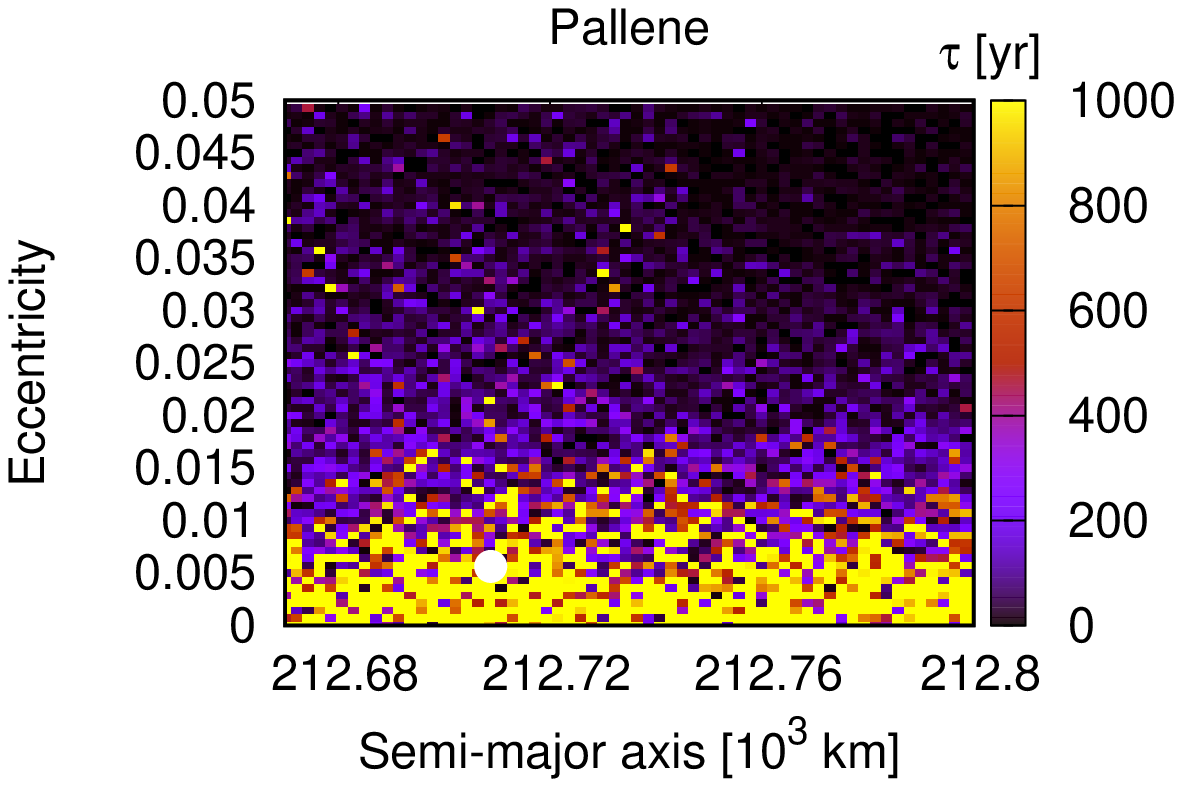}
\caption{ The same as Fig. \ref{mapas-massa} now considering that the initial angles $\omega$ and $M$ of the 4,900 test particles (in the vicinity of each small moon) are randomly distributed between $0^{\circ}$ and $360^{\circ}$ (the initial $i$ and $\Omega$ are equal to zero). The filled circles represent the values of $a$ and $e$ of each small moon for the adopted epoch, that is, 2020/03/25.}
\label{mapas-massa-aleat}
\end{center}
\end{figure}

Fig. \ref{mapas-massa-aleat} shows the results for the four small moons, where, for sake of comparison, we also integrated for $10^3$ yr.
Here, instabilities also arise as a result of collisions between test particles with the corresponding small moon in each numerical simulation. 
The survival percentages are displayed in Table \ref{tabela-survivor} for the two sets of simulations, being one corresponding to Fig. \ref{mapas-massa} (non-random angles) and the other to Fig. \ref{mapas-massa-aleat} (random $\omega$ and $M$). We note that the populations of surviving particles are sensitive to the initial adopted angles. Moreover, if we compare the results of Figs. \ref{mapas-massa} and \ref{mapas-massa-aleat}, the distributions on the $(a,e)$ space of initial conditions are significantly different, mainly for Aegaeon, Methone and Anthe, that is, for those satellites currently evolving under resonant trapping with Mimas.  

It is worth emphasize that, among the resonant small moons, Methone is the most massive (see Table \ref{tabela-massas}) and also the closest to Mimas. At first, these two conditions would imply in strong perturbations on the motions of test particles within the adopted region of initial conditions. However, according to Table \ref{tabela-survivor}, the regions in the vicinity of Aegaeon and Anthe have dynamical lifetimes which are shorter than the corresponding lifetimes for Methone and Pallene. 


\begin{table}
\begin{center}
\caption{Percentages of surviving test particles for two sets of numerical simulations corresponding to each small moon. In one case, the initial angles of the test particles are identical to the angles of the corresponding small moon, whereas in the other case the initial angles ($\omega$ and $M$) are randomly distributed between $0^{\circ}$ and $360^{\circ}$. A total of 4,900 test particles are considered in each numerical simulation over $10^3$ yr of integration time.}
\vspace {0.2cm}
\begin{tabular}{c c c c c}
\hline
Angles &  Aegaeon  & Methone & Anthe & Pallene \\ 
\hline
\hline
Non-random & 1.7 \% & 12.2 \% & 1.3 \% & 14.6 \% \\
Random & 5.9 \% & 11.3 \% & 4.6 \%  &  11.4 \% \\
\hline
\label{tabela-survivor}
\end{tabular}
\end{center}
\end{table}





\subsection{Arcs}
\label{arcs}

The \textit{Cassini} mission has revealed the presence of arcs structures around Aegaeon, Methone and Anthe. These arcs are composed by $\mu$m to cm-size dust grains particles and extend for dozens of degrees in mean longitude, with the respective satellite residing within each arc \citep{hedman+2007,hedman+2009}. Previous works suggested that the 7/6, 14/15 and 10/11 corotation resonances act as a mechanism to confine the arcs particles of Aegaeon, Methone and Anthe, respectively \citep{hedman+2007,hedman+2009}. The origin of the arcs is unclear, however, they can arise as the result of collisions of micrometeoroids with the resonant small moons. As we mentioned in Sec. \ref{clumps}, previous works have investigated the formation and sustainability of arcs structures around Aegaeon, Methone and Anthe considering gravitational and non-gravitational perturbations \citep{araujo+2016,sun+2017,madeira+2018}. Note that \cite{madeira+winter2020} claim that, due to instabilities of the dust particles, the arcs tend to be transient structures.

\begin{figure}
\begin{center}
\includegraphics[width=0.5\columnwidth,angle=270]{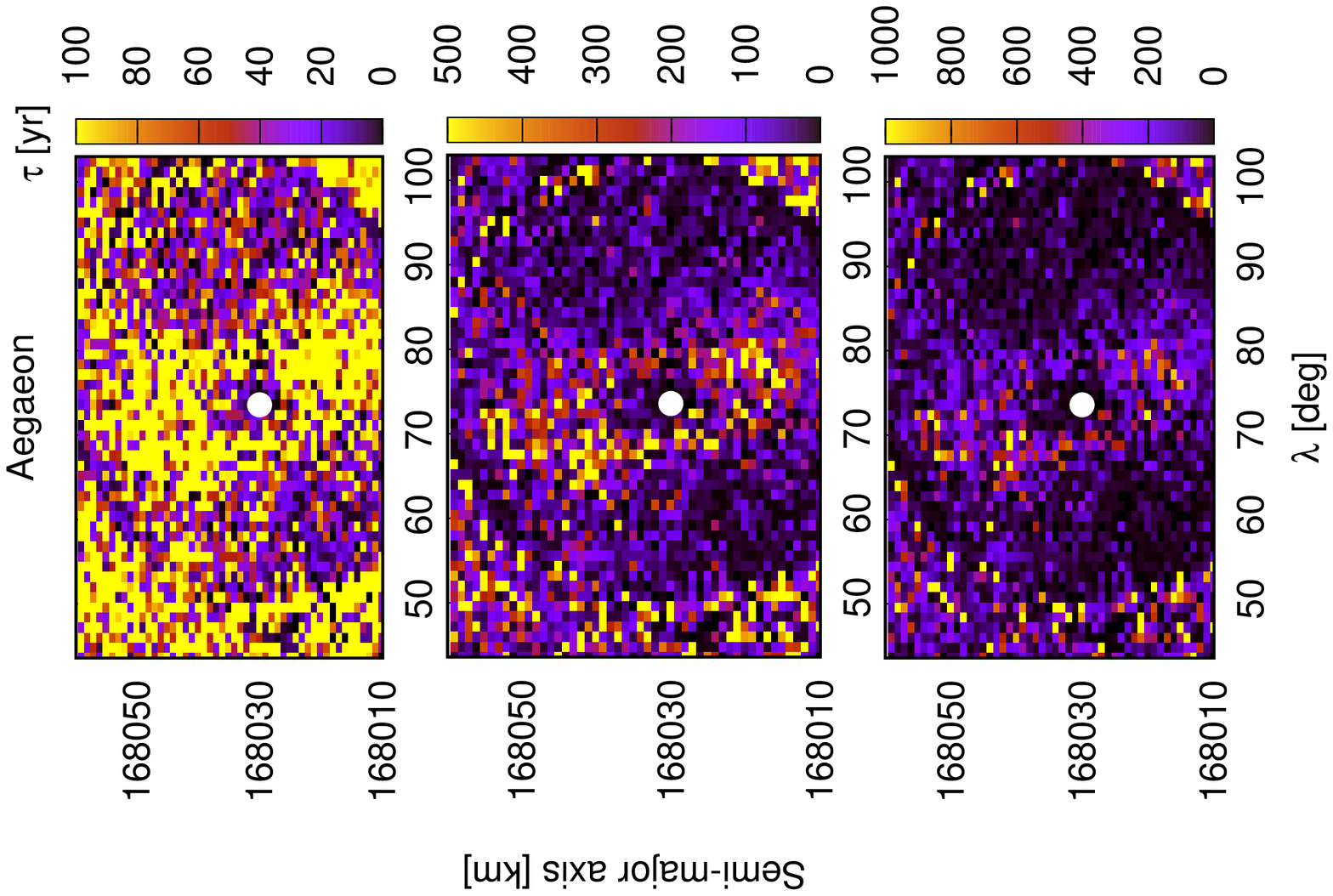}
\includegraphics[width=0.5\columnwidth,angle=270]{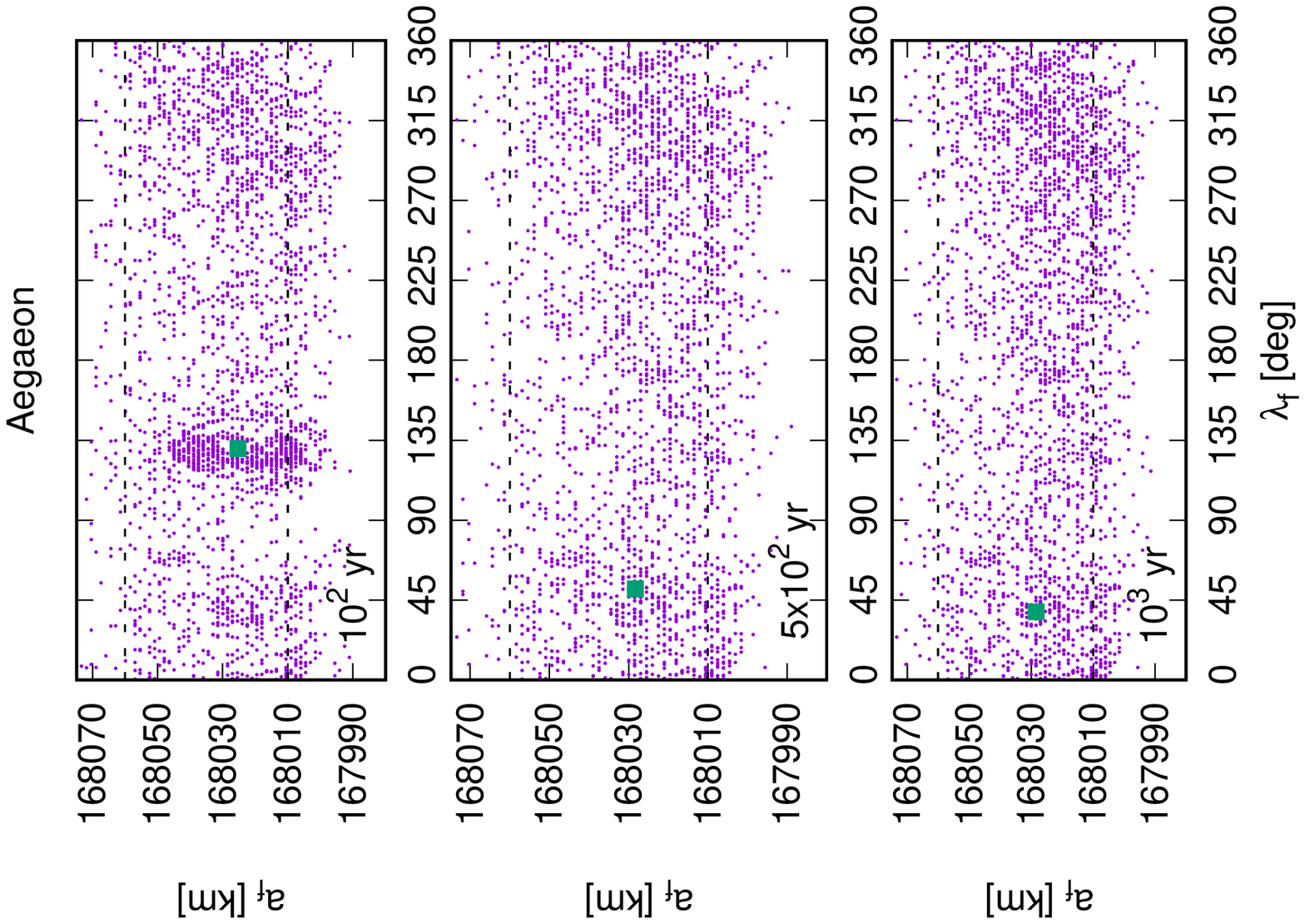}
\caption{\textit{Left}: stability time (in units of years) for 2,500 test massless particles initially distributed along the observed longitudinal extension of the arc of Aegaeon ($60^{\circ}$), with initial semi-major axes within the extension of the corotation resonance with Mimas. Results are shown for three numerical simulations with different integration times, namely, $10^2$ yr (top), $5\times10^2$ yr (middle) and $10^3$ yr (bottom). The filled dot indicates the values of Aegaeon for the adopted date (2020/03/25). \textit{Right}: Final values of mean longitudes and semi-major axes for the same three runs. The horizontal dashed lines indicate the size of the corotation region (in km), where the test particles are initially distributed. The final values for Aegaeon are indicated by green squares. The mass of Aegaeon is derived from \citet{thomas+2013}}
\label{mapas-massa-arcos-aeg}
\end{center}
\end{figure}

\begin{figure}
\begin{center}
\includegraphics[width=0.5\columnwidth,angle=270]{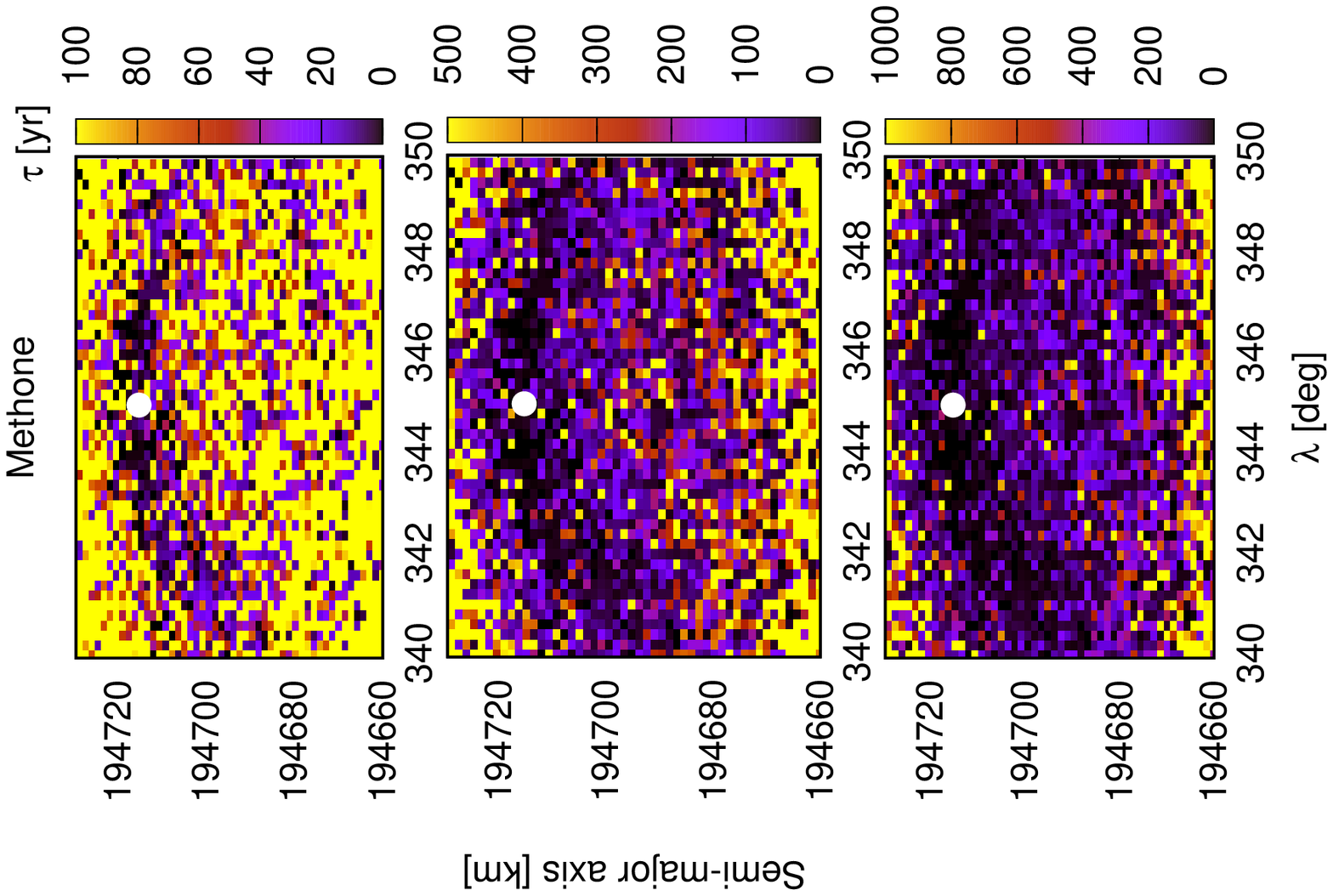}
\includegraphics[width=0.5\columnwidth,angle=270]{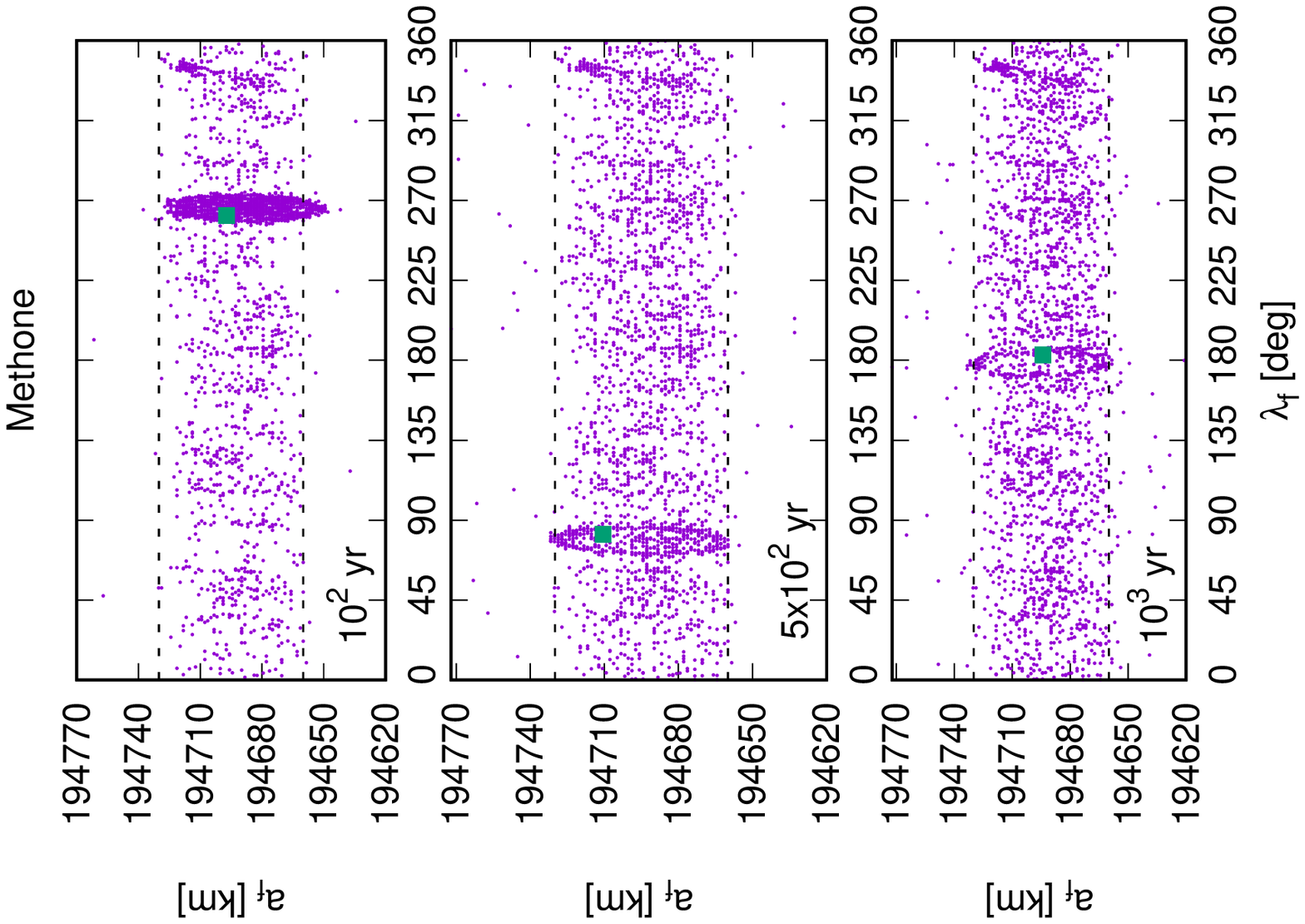}
\caption{ The same than Fig. \ref{mapas-massa-arcos-aeg} here for the arc of Methone. Note as the arc survives for the three adopted integration times.}
\label{mapas-massa-arcos-meth}
\end{center}
\end{figure}

\begin{figure}
\begin{center}
\includegraphics[width=0.5\columnwidth,angle=270]{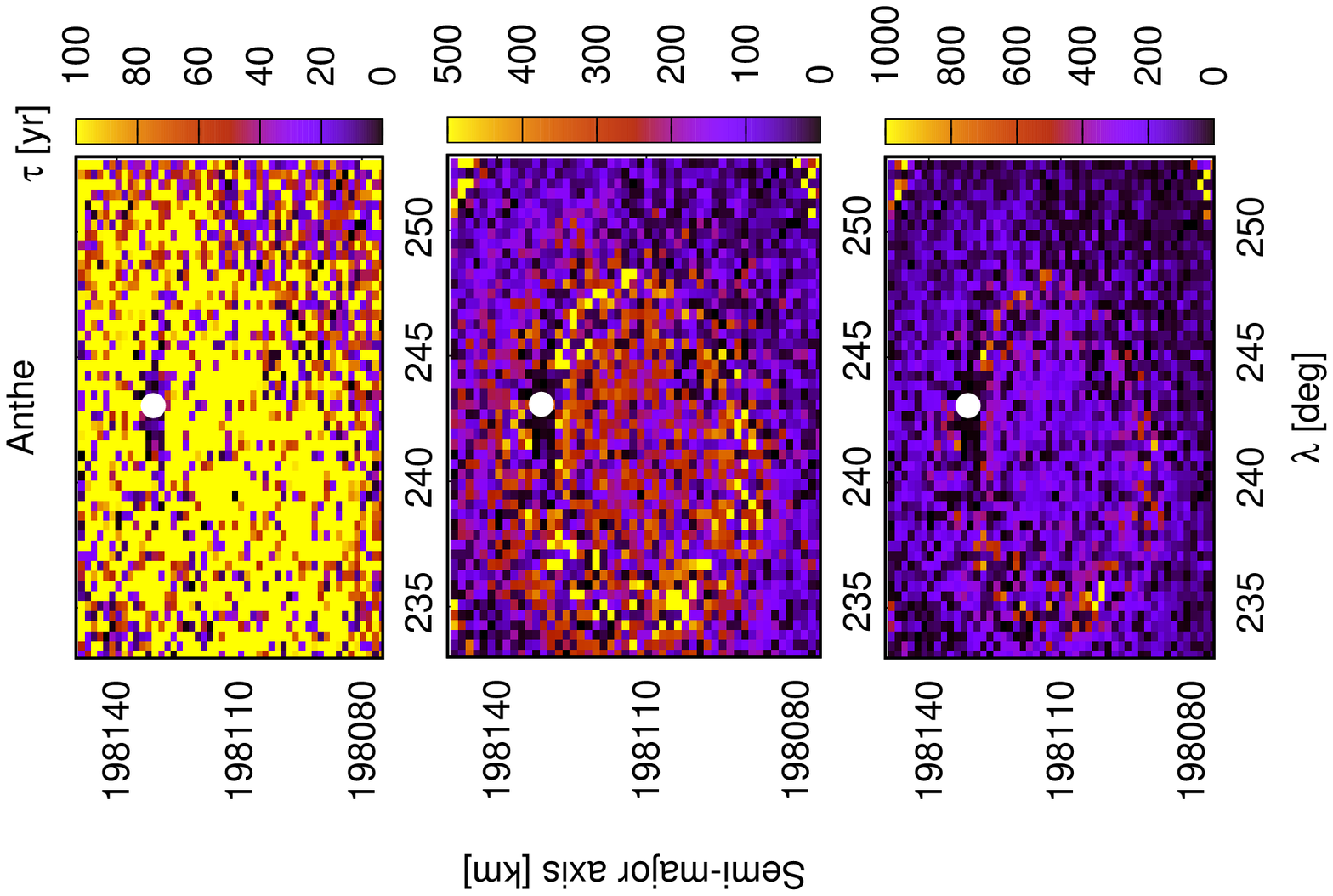}
\includegraphics[width=0.5\columnwidth,angle=270]{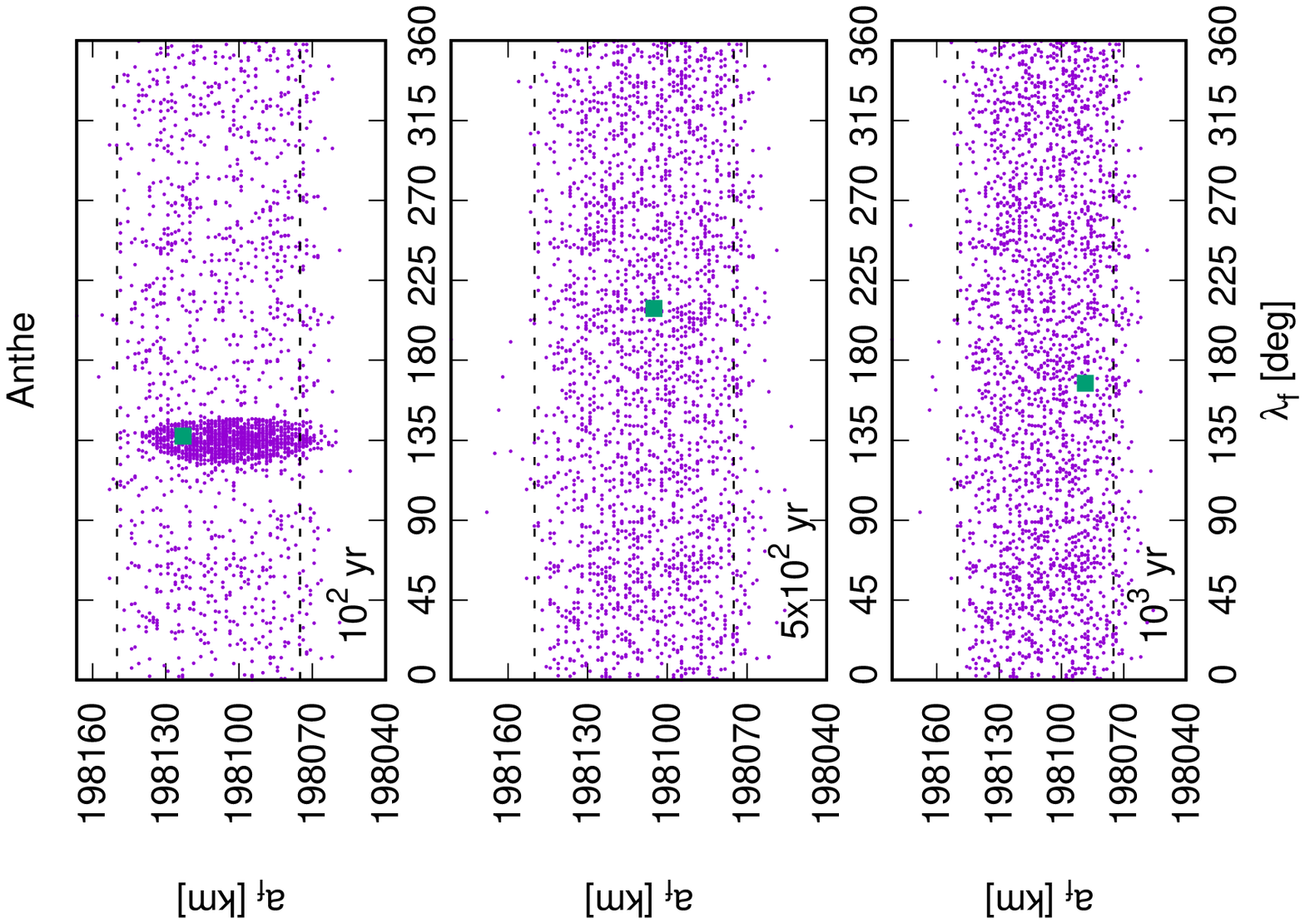}
\caption{ The same than Figs. \ref{mapas-massa-arcos-aeg} and \ref{mapas-massa-arcos-meth} here for the arc of Anthe.}
\label{mapas-massa-arcos-ant}
\end{center}
\end{figure}

In this section we aim to estimate the dynamical timescale for massless test particles undergoing only gravitational perturbation. The main goal is to highlight the influence  of gravitational perturbations, allowing us to estimate the timescale for which non-gravitational forces can operate (that is, radiation pressure, magnetic field and plasma drag). Hence, we consider 2,500 test particles varying the initial mean longitude and semi-major axis according to the longitudinal extent of each arc and covering the whole amplitude, in semi-major axis, of the corresponding CER widths (see Sec. \ref{massless}). We assume that the initial eccentricities and inclinations of test particles are the same that for each small moon \citep[see also][]{madeira+2018}. Figs. \ref{mapas-massa-arcos-aeg} to \ref{mapas-massa-arcos-ant} show the results for the stability time ($\tau$) (left panel) and also for the final distribution of mean longitudes and semi-major axes (right panel) for the arcs of Aegaeon, Methone and Anthe. We simulate for $10^2$ yr, $5\times10^2$ yr and $10^3$ yr, allowing us to better distinguish the evolution of the arcs particles for different timescales. 

According to Fig. \ref{mapas-massa-arcos-aeg}, the arc of Aegaeon survives as long as $10^2$ yr (note the clump of particles in the top right panel) and extends $\sim30^{\circ}$ in longitude. For longer timescales, the particles are removed as a result of collisions with Aegaeon. Note that \citet{madeira+2018} have shown that 75 \% of initially 6,000 particles in the arc of Aegaeon are removed within $5\times10^2$ yr of evolution under the gravitational perturbation of Aegaeon, Mimas, Tethys and Saturn oblateness.

The dynamical environment of the Methone arc reveals a different scenario as can be seen in Fig. \ref{mapas-massa-arcos-meth}. Note as the arc is dynamically eroded, however, a substantial amount of the original set of particles survives for $10^3$ yr. It is worth to note that the longitudinal extent of the arc remains roughly constant ($\sim16^{\circ}$).

In the case of Anthe, Fig. \ref{mapas-massa-arcos-ant} shows that the arc is still (although hardly) visible for $5\times10^2$ yr, having a longitudinal size $\sim25^{\circ}$. Note the circle shaped structure in the plots of the stability time (middle and bottom left panel)\footnote{It is worth emphasize that in the plots of the final values of mean longitude and semi-major axes we include all the 2,500 test particles. Hence, since the stability time depends on the specific particle, these plots cannot be considered as snapshots at $10^2$ yr, $5\times10^2$ yr and $10^3$ yr. However, as all particles belonging to the clumps have the same value of $\tau$, these clumps represent the corresponding arcs at $10^2$ yr, $5\times10^2$ yr and $10^3$ yr.}.

Table \ref{tabela-survivor-arcos} summarizes the resulting survival times for all arcs at different integration times.

The results above indicate that Aegaeon, Methone and Anthe are able to sustain their corresponding arcs particles for timescales of hundreds of years when only gravitational interactions are taken into account. Moreover, we obtained arcs with longitudinal sizes similar to the observed ones, at least for Methone and Anthe. When also non-gravitational forces are considered the lifetime of the particles is size dependent, with grains larger than 5 $\mu$m surviving in the arcs of Methone and Anthe for $10^2$ yr  \citep{sun+2017}, whereas particles of 10 $\mu$m have a lifetime of 26 yr in the arc of Aegaeon \citep{madeira+2018}. These results together put constraints on the timescales for which gravitational and non-gravitational forces operate to remove particles from the arcs.  

\begin{table}
\begin{center}
\caption{Percentages of surviving test particles initially distributed in form of arc around Aegaeon, Methone and Anthe for three different integration times. The initial arc covers the longitudinal extent of each real arc, according to the observations.}
\vspace {0.2cm}
\begin{tabular}{c c c c }
\hline
Time ($10^2$ yr) &  Aegaeon  & Methone & Anthe \\ 
\hline
\hline
1 & 35.4 \% & 48.2 \% & 53.6 \%  \\
5 & 5.8 \% & 17 \% & 2.8 \%  \\
10 & 2.0 \% & 9.7 \% & 0.36 \%  \\
\hline
\label{tabela-survivor-arcos}
\end{tabular}
\end{center}
\end{table}

\section{Discussion and conclusions}\label{discussion}

In this work we analyze the orbital evolution and the dynamical stability in the vicinity of Aegaeon, Methone, Anthe and Pallene. Through a large set of numerical simulations we investigate the orbital motion of test particles within and near to the domain of the 7/6, 14/15, 10/11 mean motion resonances of Aegaeon, Methone and Anthe with Mimas, respectively. Our model includes the gravitational perturbation of companion satellites from Mimas to Titan and the contribution of the Saturn's oblateness quantified by the $J_2$ and $J_4$ coefficients of the central potential.

On one hand, we have shown that when each small satellite is considered as a massless body, all initial particles restricted to resonant motions remain stable for at least $10^4$ yr. In the cases of Methone and Anthe a significant amount of test particles initially out of resonance undergo large variations in semi-major axis, eccentricity and inclination, ultimately resulting in collisions with Mimas in timescales of a few thousands of years (see Figs. \ref{mapa-methone} and \ref{mapa-anthe}). For  Aegaeon, all particles remain in stable orbits within the adopted ranges of initial semi-major axis and eccentricity, revealing a dynamical environment of regular motion. Moreover, Aegaeon is located very close to the centre of the 7/6 CER with Mimas (see Fig. \ref{mapa-aegaeon}). The structures associated to the corotation resonances of Methone, Anthe and Aegaeon in the space of semi-major axis and eccentricity (Figs. \ref{mapa-methone}, \ref{mapa-anthe} and \ref{mapa-aegaeon}) have maximum widths for very small values of the eccentricity. In addition, the widths decreases as the eccentricity increases resulting in inverted V-shape structures (see Callegari et al. 2021, submitted). The vicinity of Pallene is also characterized by small amplitude variations of the orbital elements of test particles, mainly for initial eccentricities smaller than 0.03 (see Fig. \ref{mapa-pallene}). We have also shown the existence of clumps in the final distribution of semi-major axes and mean longitudes of test particles (see Fig. \ref{arcos1}). These clumps are the natural consequence of the confinement of particles due to the trapping in the corresponding corotation resonance (see Eq. \ref{eq-arcos}; see also \citet{hedman+2009}).

On the other hand, we also consider that Aegaeon, Methone, Anthe and Pallene are massive bodies in our numerical simulations. The masses are derived from the radii and mean densities reported in \citet{thomas+2013}. The results shown that the perturbation of the small moons have a severe effect on the motion of surrounding particles, destabilizing most of them in a timescale of a few hundreds of years or even less (see Figs. \ref{mapas-massa} and \ref{mapas-massa-aleat}). {When the initial mean longitudes of the test particles are considered as the same of the mean longitudes of the small moon, the survival percentage is smaller than for the case of initial random mean longitudes. This fact highlights the importance of the initial mean longitudes for the dynamical behaviour of particles in corotation resonances. In addition, we have found that all unstable particles collide with the corresponding small moon within the adopted timescale.}

We also simulated the orbital behaviour of test particles initially spread in form of arc around Aegaeon, Methone and Anthe, evolving only under gravitational perturbation. We have shown that the initial arcs are dynamically eroded on timescales of hundreds of years (see Figs. \ref{mapas-massa-arcos-aeg}\,-\,\ref{mapas-massa-arcos-ant}), putting restrictions on the timescales for which gravitational and non-gravitational forces operate to remove particles from the arcs. {Note that when initial random angles are considered, the stability of particles seems to increase (compare Figs. \ref{mapas-massa-aleat} and \ref{mapas-massa-arcos-aeg} to \ref{mapas-massa-arcos-ant}), indicating that other corotation resonance sites which are not populated by dust particles would be more stables than the observed arcs. This fact reinforce the hypothesis that the small moons act as the source of the observed arcs.}

{We have reproduced the longitudinal extent of the Methone and Anthe arcs in agreement with the observed arcs. However, for the case of Aegaeon, the simulated arc seems to be moderately smaller than the observed one (see Figs. \ref{mapas-massa-arcos-aeg} to \ref{mapas-massa-arcos-ant}). Perhaps, additional effects would be considered in the analysis for the case of Aegaeon. It is worth to mentioning that we are investigating the 7/6 Aegaeon - Mimas corotation resonance in more detail in a forthcoming paper.}

Within the limitations of our model we can cite:  \textit{i}) the non consideration of additional sources of gravitational perturbation such as the Sun, companion planets and satellites (i.e, Janus and Epimetheus). \textit{ii}) The absence of the rings. We believe that these two limitations would not significantly affect the stability analysis presented in this work, however, the long-term orbital motion of the small moons (specially Aegaeon) can be modified. We will address the long-term stability of Aegaeon, Methone, Anthe and Pallene in a future work, including the above mentioned contributions. In addition, Mimas and other saturnian mid-sized satellites have experienced orbital migration due to tidal interaction with Saturn \citep[see][]{nakajima+2019,nakajima+2020}. The investigation of the orbital stability of the small moons and their arc particles, coupled with the Mimas tidal migration, would allows us to better understand the recent dynamics of corotation resonances \citep[e.g.][]{araujo+2016}, as well as to point out possible origins for these resonant motions. We will also address these points in a future work.

\section*{Acknowledgements}

We acknowledge the anonymous reviewer for his/her very detailed revision and for the helpful suggestions, which certainly allowed us to improve the new version of the manuscript. We specially thanks to Fernando Roig for a critical reading and suggestions. The $N$-body numerical simulations have been performed in the Lobo Carneiro supercomputer from NACAD\,-\,Coppe, Universidade Federal do Rio de Janeiro (UFRJ). NCJ thanks to Fapesp, process 2020/06807-7.

\section*{Data Availability}

The data underlying this article will be shared on reasonable request to the corresponding author.



\bibliographystyle{mnras}
\bibliography{references} 



\appendix

\section{Graphics using the same scaling}

\begin{figure}
\begin{center}
\includegraphics[width=0.45\columnwidth,angle=270]{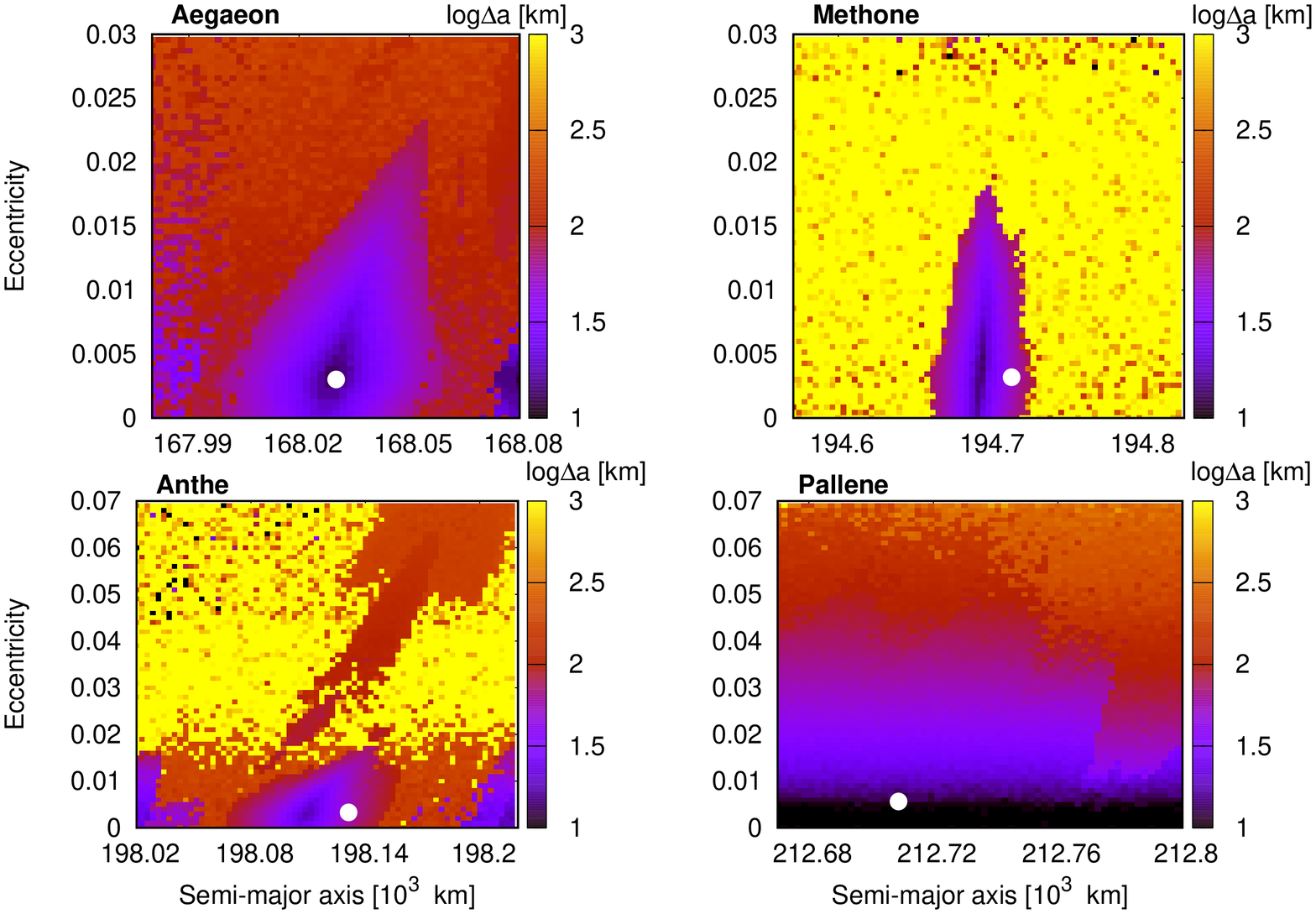}
\caption{Comparison of maximum variations of semi-major axis for the four small satellites according to the simulations described in Sec. \ref{massless}. Note that here all panels are shown using the same color scale.}
\label{mapa-semi}
\end{center}
\end{figure}

\begin{figure}
\begin{center}
\includegraphics[width=0.45\columnwidth,angle=270]{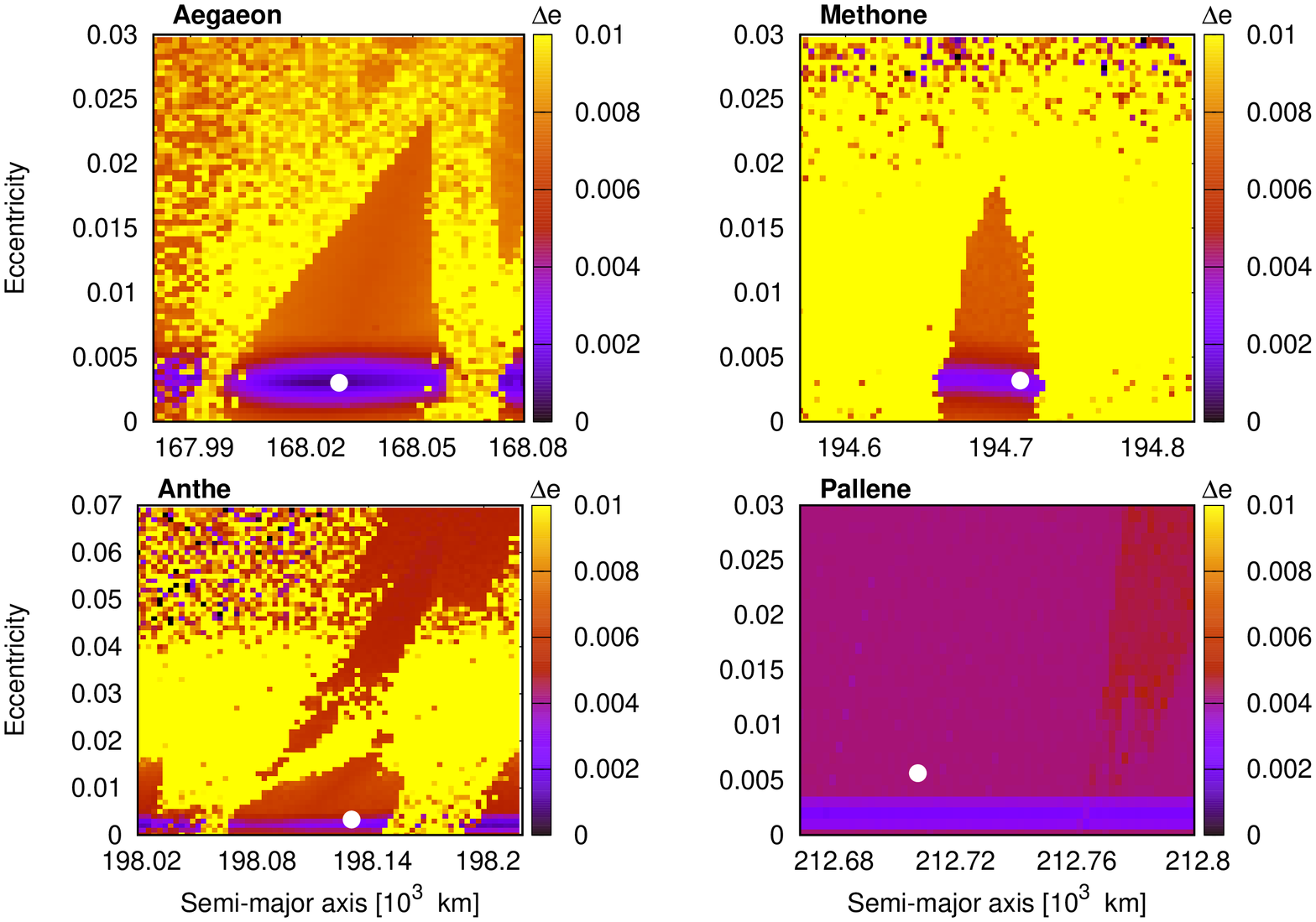}
\caption{The same of Fig. \ref{mapa-semi} now showing the maximum variations in orbital eccentricity.}
\label{mapa-exce}
\end{center}
\end{figure}

\begin{figure}
\begin{center}
\includegraphics[width=0.45\columnwidth,angle=270]{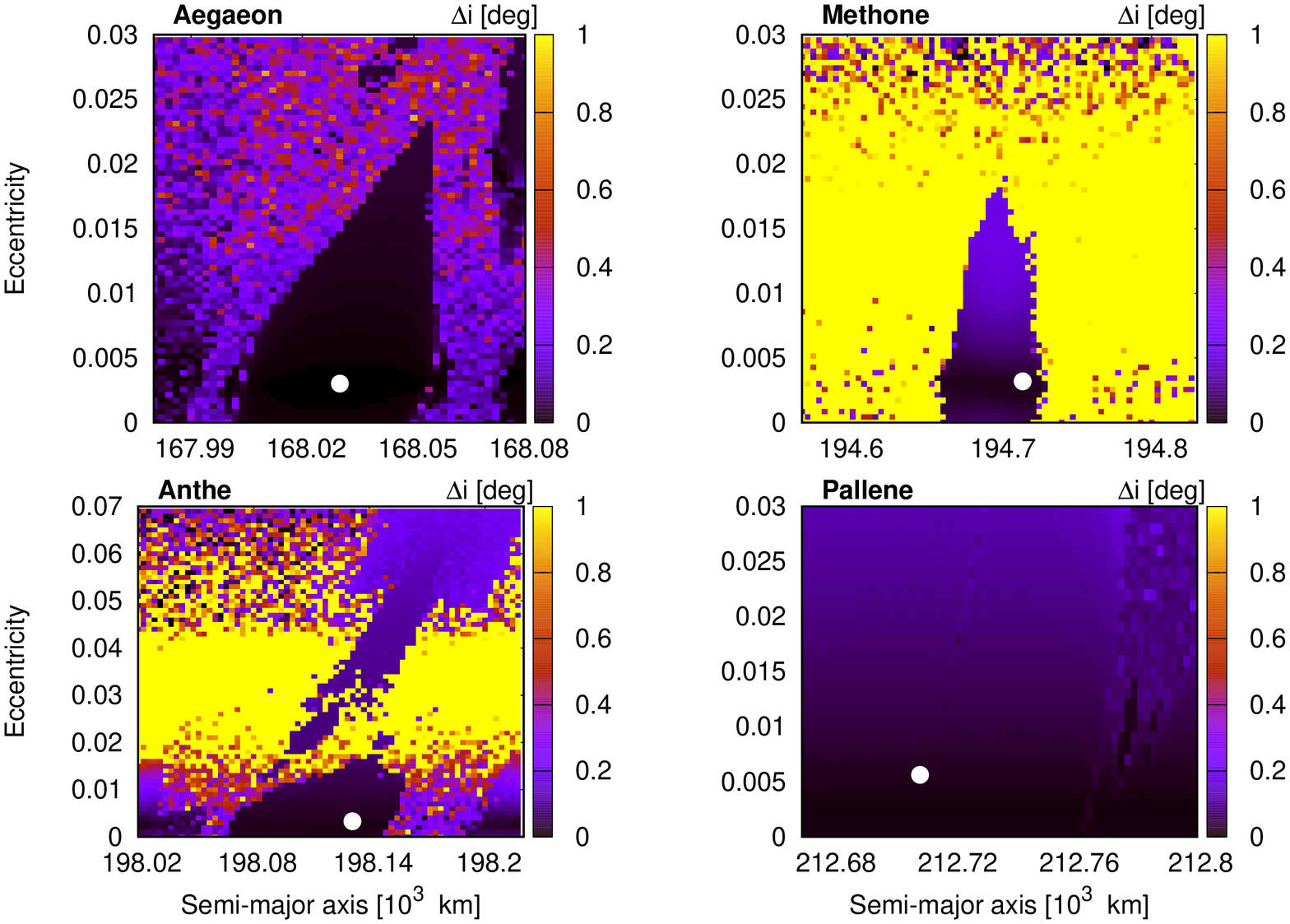}
\caption{The same of Fig. \ref{mapa-semi} now showing the maximum variations in orbital inclination.}
\label{mapa-incli}
\end{center}
\end{figure}

We show in Figs. \ref{mapa-semi}, \ref{mapa-exce} and \ref{mapa-incli} the same results presented in Figs. \ref{mapa-methone}, \ref{mapa-anthe} and \ref{mapa-aegaeon} but now considering the same color scale and grouping in maximum variations of orbital elements ($\Delta a$, $\Delta e$ and $\Delta i$). This grouping allows us to better compare the variations for the four satellites.

\bsp	
\label{lastpage}
\end{document}